\documentclass{aa}  

\usepackage{graphicx}
\usepackage{txfonts}
\begin{document} 

\title{{\it BRITE} photometry of the massive post-RLOF system HD~149\,404\thanks{Based on data collected by the {\it BRITE}-Constellation satellite mission, designed, built, launched, operated and supported by the Austrian Research Promotion Agency (FFG), the University of Vienna, the Technical University of Graz, the University of Innsbruck, the Canadian Space Agency (CSA), the University of Toronto Institute for Aerospace Studies (UTIAS), the Foundation for Polish Science \& Technology (FNiTP MNiSW), and the National Science Centre (NCN).}}
\author{G.\ Rauw\inst{1} \and A.\ Pigulski\inst{2} \and Y.\ Naz\'e\inst{1}\fnmsep\thanks{Research Associate FRS-FNRS (Belgium).} \and A.\ David-Uraz\inst{3} \and G.\ Handler\inst{4} \and F.\ Raucq\inst{1} \and E.\ Gosset\inst{1}\fnmsep\thanks{Research Director FRS-FNRS (Belgium)} \and A.F.J.\ Moffat\inst{5} \and \\ C.\ Neiner\inst{6} \and H.\ Pablo\inst{7} \and A.\ Popowicz\inst{8} \and S.M.\ Rucinski\inst{9} \and G.A.\ Wade\inst{10} \and W.\ Weiss\inst{11} \and K.\ Zwintz\inst{12}}
\institute{Space sciences, Technologies and Astrophysics Research (STAR) Institute, Universit\'e de Li\`ege, All\'ee du 6 Ao\^ut, 19c, B\^at B5c, 4000 Li\`ege, Belgium\\ \email{g.rauw@uliege.be} \and Instytut Astronomiczny, Uniwersytet Wroc\l awski, Kopernika 11, 51-622 Wroc\l aw, Poland \and Department of Physics and Astronomy, University of Delaware, Newark, DE 19716, USA \and Nicolaus Copernicus Astronomical Center, Bartycka 18, 00-716 Warszawa, Poland \and  D\'epartement de Physique, Universit\'e de Montr\'eal, C.P.\ 6128, Succ.\ Centre-Ville, Montr\'eal, PQ H3C 3J7, Canada \and LESIA, Observatoire de Paris, Place Jules Janssen, 5, 92195 Meudon, France \and AAVSO Headquarters, 49 Bay State Rd., Cambridge, MA 02138, USA \and Instytut Automatyki, Wydzia\l Automatyki Elektroniki i Informatyki, Politechnika \'{S}l\k{a}ska, Akademicka 16, 44-100 Gliwice, Poland \and Department of Physics and Space Science, Royal Military College of Canada, PO Box 17000, Stn Forces, Kingston, Ontario K7K 7B4, Canada \and Department of Astronomy and Astrophysics, University of Toronto, 50 St.\ George St., Toronto, Ontario, M5S 3H4, Canada \and Institute for Astrophysics, University of Vienna, T\"urkenschanzstra\ss e 17, 1180 Vienna, Austria \and Institute for Astro- and Particle Physics, University of Innsbruck, Technikerstrasse 25/8, 6020 Innsbruck, Austria} 
\date{}

 
  \abstract
   {HD~149\,404 is an evolved non-eclipsing O-star binary that has previously undergone a Roche lobe overflow interaction.}{Understanding some key properties of the system requires a determination of the orbital inclination and of the dimensions of the components.}{The {\it BRITE-Heweliusz} satellite was used to collect photometric data of HD~149\,404. Additional photometry was retrieved from the SMEI archive. These data were analysed using a suite of period search tools. The orbital part of the lightcurve was modelled with the {\tt nightfall} binary star code. The {\it Gaia}-DR2 parallax of HD~149\,404 was used to provide additional constraints.}{The periodograms reveal a clear orbital modulation of the lightcurve with a peak-to-peak amplitude near 0.04\,mag. The remaining non-orbital part of the variability is consistent with red noise. The lightcurve folded with the orbital period reveals ellipsoidal variations, but no eclipses. The minimum when the secondary star is in inferior conjunction is deeper than the other minimum due to mutual reflection effects between the stars. Combined with the {\it Gaia}-DR2 parallaxes, the photometric data indicate an orbital inclination in the range of $23^{\circ}$ to $31^{\circ}$ and a Roche lobe filling factor of the secondary larger than or equal to 0.96.}{The luminosity of the primary star is consistent with its present-day mass, whereas the more evolved secondary appears overluminous for its mass. We confirm  that the primary's rotation period is about half the orbital period. Both features most probably stem from the past Roche lobe overflow episode.}
\maketitle
%

\section{Introduction}
It has been suspected for quite a long time that HD~149\,404 \citep[$\equiv $V~918 Sco; O7.5\,I(f) + ON9.7\,I, $e=0.0$, $P_{\rm orb} = 9.81$\,days,][see also the parameters in Table\,\ref{param}]{Rauw} is a post-Roche lobe overflow (RLOF) binary system  \citep[e.g.][]{Penny}. Quite recently, this scenario was confirmed via an in-depth analysis of its optical spectrum \citep{Raucq}. Following spectral disentangling and modelling of the spectrum with the CMFGEN model atmosphere code \citep{CMFGEN}, \citet{Raucq} established a nitrogen overabundance with [N/C] $\sim 150$\,[N/C]$_{\odot}$ for the present-day secondary star\footnote{In accordance with previous work on HD~149\,404, we call the primary (resp.\ secondary) the currently more (resp.\ less) massive component.}, which can best be explained if the system has gone through an episode of Case A Roche lobe overflow. These properties make HD~149\,404 an excellent candidate for an O-star binary on its way to becoming a Wolf-Rayet binary via the binary evolution channel. Yet, there remain several open issues in this scenario that are directly related to the uncertain orbital inclination.

The most important issue concerns the masses of the components: are they typical for the spectral types of these stars, or are they significantly lower, as in the case of LSS~3074, another post-RLOF early-type binary \citep{Raucq2}? The orbital solution of the system \citep{Rauw} yielded very low minimum masses $m_1\,\sin^3{i} = 2.52$\,M$_{\odot}$ and $m_2\,\sin^3{i} = 1.52$\,M$_{\odot}$ (see Table\,\ref{param}). Without a knowledge of the orbital inclination, it is impossible to tell whether these numbers reflect unusually low stellar masses or just a low orbital inclination. Another uncertainty regarding the evolution of HD~149\,404 is the transfer of angular momentum. An efficient exchange of angular momentum in a Roche lobe overflow interaction could speed up the rotation of the mass gainer to the point that it repels any further mass coming from the mass loser \citep{Petrovic}. A good diagnostic for such an angular momentum transfer is a strong asynchronicity of the rotation of the stars. Whilst there are some hints for asynchronous rotation in HD~149\,404 \citep{Raucq}, the conclusion depends again on the orbital inclination.

\begin{table*}[t!]
  \caption{Important parameters of HD~149\,404\label{param}}
  \begin{center}
    \begin{tabular}{c c c c c}
      \hline
      & Primary & Secondary & Binary & Ref.\\
      \hline
      $P_{\rm orb}$ (days) & & & $9.81475 \pm 0.00084$ & (1) \\
      $T_0$ (HJD$-$2\,450\,000) & & & $1680.279 \pm 0.174$ & (1) \\
      $a\,\sin{i}$ (R$_{\odot}$) & & & $30.7 \pm 0.7$ & (1) \\ 
      $R_{\rm RL}/a$ & $0.42 \pm 0.01$ & $0.34 \pm 0.01$ & & (1) \\
      $m\,\sin^3{i}$ (M$_{\odot}$) & $2.52 \pm 0.21$ & $1.52 \pm 0.13$ & & (1) \\
      $q = m_2/m_1$ &  &  & $0.605 \pm 0.027$ &  (1) \\
      $T_{\rm eff}$ (kK) & $34.0 \pm 1.5$ & $28.0 \pm 1.5$ & & (2) \\
      $l_1/l_2$ & & & $0.70 \pm 0.12$ & (2) \\
      $BC$ (mag) & $-3.17$ & $-2.67$ & & (2) \\
      $v\,\sin{i}$ (km\,s$^{-1}$) & $93 \pm 8$ & $63 \pm 8$ & & (2) \\
      \hline
    \end{tabular}
  \end{center}
  \tablefoot{$R_{\rm RL}$, $m$, and $BC$ are the equivalent radius of the component's Roche lobe, the mass of the star, and its bolometric correction in the $V$-band, respectively. The orbital separation of the system is given by $a$, and  the optical brightness ratio of the primary star  to the secondary is $l_1/l_2$. The references are (1) \citet{Rauw} and (2) \citet{Raucq}.}
\end{table*}

Our ignorance about the orbital inclination stems from the fact that HD~149\,404 does not display genuine photometric eclipses, though Dr.\ N.\ Morrison found low-level photometric variability in ground-based observations \citep{MC}. \citet{Luna} collected polarimetric observations of the system that led to an upper limit on the orbital inclination of $50^{\circ}$. Comparing the minimum masses of the stars inferred from the orbital solution (see Table\,\ref{param}) to typical spectroscopic masses of stars of the same spectral type, \citet{Rauw} inferred an inclination near $21^{\circ}$. However, given the peculiar evolutionary state of the system, such a comparison and hence the resulting value of $i$ are subject to major uncertainties. Whilst ground-based observations did not show clear orbital variations \citep{MC}, higher-sensitivity space-borne photometry has the potential to reveal low-level variability possibly resulting from ellipsoidal variations \citep[see e.g.\ the case of Plaskett's Star, HD~47\,129,][]{Mahy}.

Such orbital variations can also help us infer the dimensions of the stars, which are notably important for interpreting the interactions in the system. The H$\alpha$ and He\,{\sc ii} $\lambda$\,4686 emission lines in the spectrum of HD~149\,404 undergo a strong phase-locked line profile variability \citep{Rauw,Thaller,Naze}. These lines notably exhibit an unusual double-peaked morphology at orbital phases close to conjunction. The line-profile variations were interpreted either as the signature of a wind-wind interaction with an outer interaction zone (where the optical emission lines are formed) that is strongly bent by the Coriolis pseudo-forces \citep{Rauw,Naze} or as the result of the focused winds flowing into the wind interaction zone \citep{Thaller}. 

In this context, we report here the results of a photometric monitoring campaign of HD~149\,404 with the {\it Heweliusz} satellite of the {\it BRITE} constellation \citep{Weiss} combined with the analysis of archival SMEI photometry \citep{SMEI}. Section\,\ref{obs} presents the observational data used in this work along with the data reduction. The frequency content of the lightcurves as well as the properties of the non-orbital variability are analysed in Sect.\,\ref{lcBRITE}. Section\,\ref{modelorb} deals with the analysis and interpretation of the orbital variability. Finally, the implications of our results are discussed in Sect.\,\ref{discuss} and the conclusions are presented in Sect.\,\ref{conclusion}.  

\section{Observations and data processing \label{obs}}
\subsection{{\it BRITE-Heweliusz} data}
HD~149\,404 was observed by the {\it BRITE-Heweliusz} satellite in the framework of the Scorpius I field observation between 26 June 2015 and 28 August 2015. This satellite is part of the BRIght Target Explorer ({\it BRITE}) constellation consisting of five nano-satellites observing through blue and red filters. {\it BRITE} is operated by Austrian, Canadian, and Polish institutes and its goal is   collecting extended time series of high-precision photometry of stars brighter than $V \simeq 6$\,mag in an effective field of view of $24^{\circ} \times 20^{\circ}$ \citep[see][for a detailed description]{Weiss,Pablo}. {\it BRITE-Heweliusz} uses a four-lens 3~cm aperture telescope combined with a red filter with a passband extending from $\sim 5500$ to $\sim 7000$\,\AA. In the HD~149\,404 campaign, a total of 21\,448 data points were taken with exposure times of 2.5\,s. These observations were performed in chopping mode \citep{Pablo}. 

\begin{figure}[h]
\begin{center}
\resizebox{8cm}{!}{\includegraphics{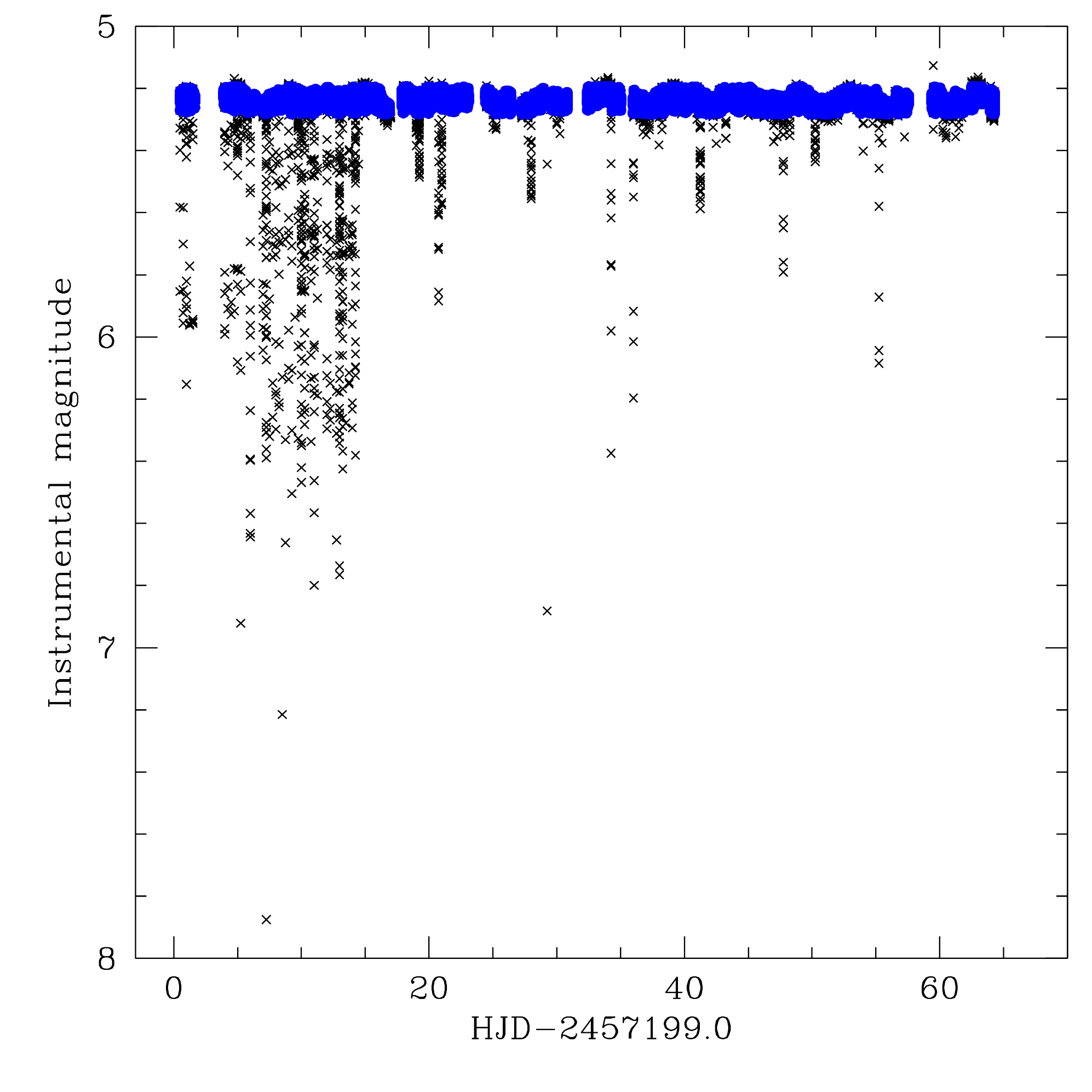}}
\end{center}  
\caption{Removal of outliers from the {\it BRITE} photometry of HD~149\,404. The black symbols indicate the raw data, the blue symbols stand for the results after MAD filtering and an additional $3\,\sigma$ clipping around the mean magnitude per spacecraft orbit.\label{outliers}}
\end{figure}

The raw data were extracted with the dedicated pipeline \citep{Popowicz}. A periodogram of the raw photometry was computed with the method of \citet{HMM} and \citet{Gosset} that explicitly accounts for uneven sampling. The strongest peak is found at the orbital frequency (14.827\,d$^{-1}$) of the {\it BRITE-Heweliusz} satellite. This is due to extreme outliers appearing in the raw data. Additional cleaning of the data was therefore required. 

To remove outliers, we first applied a median absolute deviation (MAD) filtering, discarding values that deviate by more than
\begin{equation}
  3 \times {\rm median}(|X_i - {\rm median}{(X_i)}|)
,\end{equation}
where $X_i$ is the instrumental magnitude of HD~149\,404. In this way, we mostly removed points with very low fluxes. As a second cleaning step, we then performed a $3\,\sigma$ clipping around the mean magnitude per spacecraft orbit. This left us with a total of 19\,902 data points (see Fig.\,\ref{outliers}).

The Fourier periodogram of the cleaned time series shows the strongest peak at a frequency of $0.2047 \pm 0.0016$\,d$^{-1}$ (i.e.\ a period of $4.885 \pm 0.039$\,day), corresponding within the errors to half the orbital period of HD~149\,404 \citep[$P_{\rm orb}/2 = 4.9074 \pm 0.0004$\,day,][]{Rauw}\footnote{The full {\it BRITE} time series spans $\Delta\,t =  63.5$\,days. The natural width of the peaks in the periodogram is thus expected to be $\Delta\,\nu_{\rm nat} = 1/\Delta\,t = 1.57\,10^{-2}$\,d$^{-1}$. In our analysis, we conservatively assume that we can determine the position of a peak in the periodogram with an accuracy of $\Delta\,\nu_{\rm nat}/10$.}.   
\begin{figure}[h]
\begin{center}
\resizebox{8cm}{!}{\includegraphics{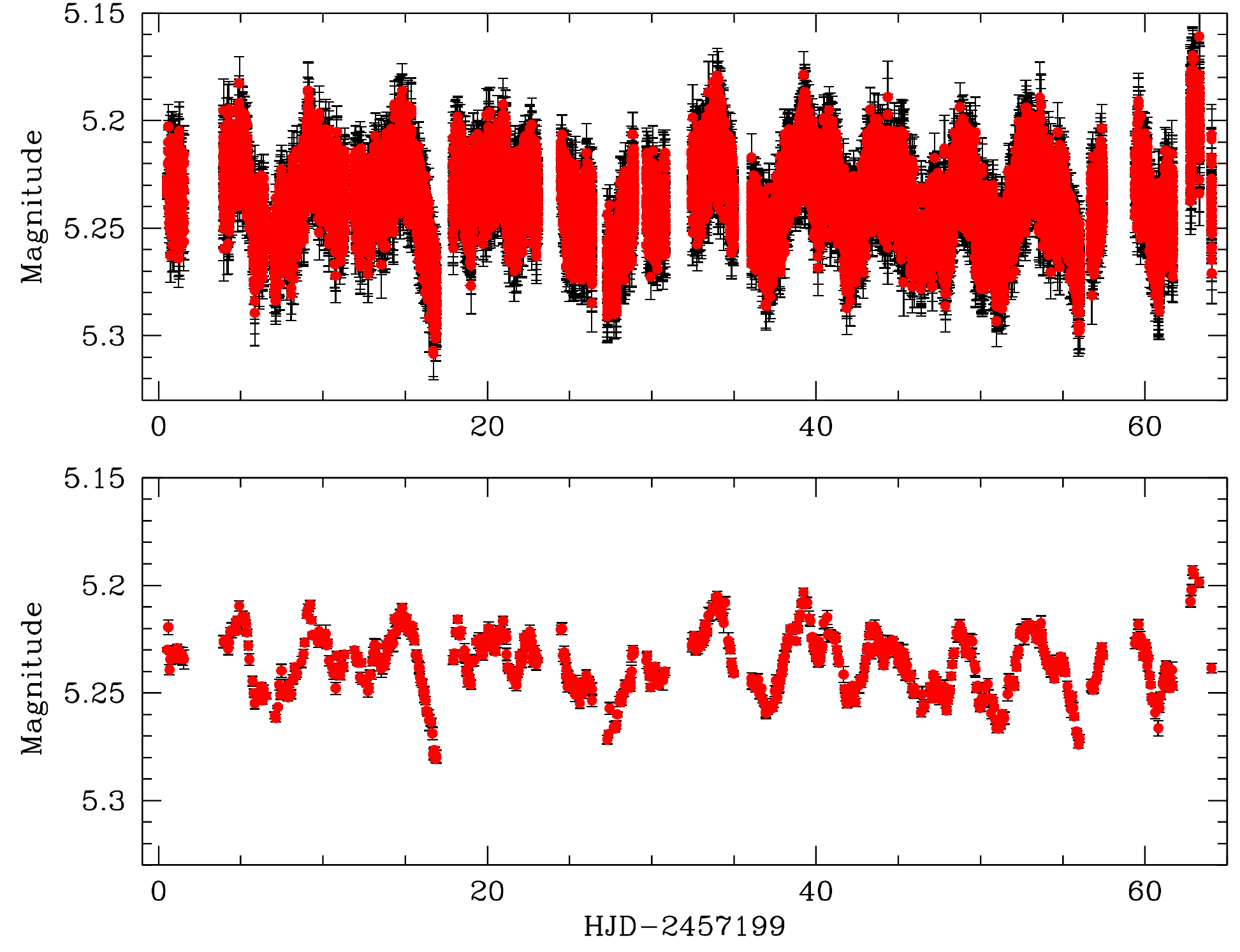}}
\end{center}
\caption{Top: Full detrended {\it BRITE} lightcurve of HD~149\,404. All 19\,414 data points are shown (red) along with their error bars (black). Bottom: Same, but for the data averaged per spacecraft orbit. This lightcurve contains 708 data points.\label{lcdetrend}}
\end{figure}

As a next step, we pre-whitened our data for the HD~149\,404 orbital frequency and twice this frequency by fitting the expression
\begin{equation}
  A_1\,\sin{(2\,\pi\,\nu_{\rm orb}\,t + \psi_1)} + A_2\,\sin{(4\,\pi\,\nu_{\rm orb}\,t + \psi_2)}
  \label{eqprew}
\end{equation}
to the data, where $A_j$ and $\psi_j$ stand for the semi-amplitude and phase constant of the sine wave of frequency $j \times \nu_{\rm orb}$. The residuals obtained in this way allowed us to check for trends with instrumental parameters such as detector temperature, $x$ and $y$ detector coordinates of the point spread function (PSF) centroid, spacecraft orbital phase, and the PSF smearing parameter. Only weak correlations were found. The strongest correlations concerned the detector temperature and $x$ detector coordinates, and they amount  to a small fraction ($< 15$\%) of the actual dispersion of the data. Decorrelation of the photometry with respect to these instrumental parameters was then performed according to the procedure outlined in the {\it BRITE} Cookbook 2.0 \citep{Pigulski18} which is an updated version of the recipe described in Appendix A of \citet{Pigulski}. In this procedure, we rejected outliers using the Generalized Extreme Studentized Deviate (GESD) algorithm described by \citet{Pigulski18}, independently of the procedure described above. For this purpose, we thus started again from the complete original data set (21\,448 data points), and first cut off the most extreme values, either in terms of the magnitude or of one of the above instrumental parameters \citep[see Sect.\ 5.2 in][]{Pigulski18}. This left us with 20\,598 data points. We then rejected the GESD outliers with a level of significance parameter $\alpha = 0.05$ \citep[see discussion on the choice of this parameter by][]{Pigulski18}. Next, we removed the spacecraft orbits containing fewer than eight data points. 
  A second GESD outlier rejection was performed with $\alpha = 0.3$ during 1-D decorrelations.
  As a next step, we removed those spacecraft orbits that have a scatter $\sigma$ higher than 0.020\,mag. 
We finally performed a third GESD outlier rejection (again with $\alpha=0.3$) after 2-D decorrelations, and rejected those spacecraft orbits for which the scatter exceeded 0.017\,mag. The full procedure resulted in a final set of 19\,414 data points. In the following, we focus on these decorrelated data. The typical error bar on the individual decorrelated magnitudes is 0.010\,mag. These error bars are computed from the scatter about the decorrelation model for data within each spacecraft orbit \citep{Popowicz}. Therefore, all data from a given spacecraft orbit are assigned the same error bar.

\subsection{SMEI data}
To complement the {\it BRITE} data, we also analysed the photometry coming from the Solar Mass Ejection Imager \citep[SMEI,][]{SMEI} experiment on board of the {\it Coriolis} satellite. {\it Coriolis} was in operation between January 2003 and September 2012. The SMEI instrument was designed for the imaging of coronal mass ejection (CME) events via the detection of solar light scattered by the free electrons in the ejected plasma. For this purpose, the instrument carried three wide field cameras that provided a nearly full sky coverage. To build maps of CME events, the light from background stars had to be subtracted. The sky was covered once per spacecraft revolution leading to a photometric measurement every 101.6\,min for each star within the accessible field of view. The PSF was about $1^{\circ}$. Therefore, nearly eight years of optical photometry of bright stars were collected as a by-product of the SMEI operations\footnote{The data processed with the SMEI pipeline \citep{Hick} are  available through the UCSD web page (http://smei.ucsd.edu/new\_smei/data\&images/stars/timeseries.html).}. The SMEI passband was defined by the response of the CCD. The latter reached a maximum quantum efficiency of $\geq 40$\% in the spectral region between 6000 and 7500\,\AA, and dropped below 10\,\% outside the spectral range from 4500 to 9500\,\AA\ \citep{Eyles}. The SMEI passband is thus considerably broader than that of the {\it BRITE-Heweliusz} satellite, although they both peak at similar wavelengths.   
\begin{figure*}[thb!]
\begin{minipage}{9cm}
\begin{center}
\resizebox{9cm}{!}{\includegraphics{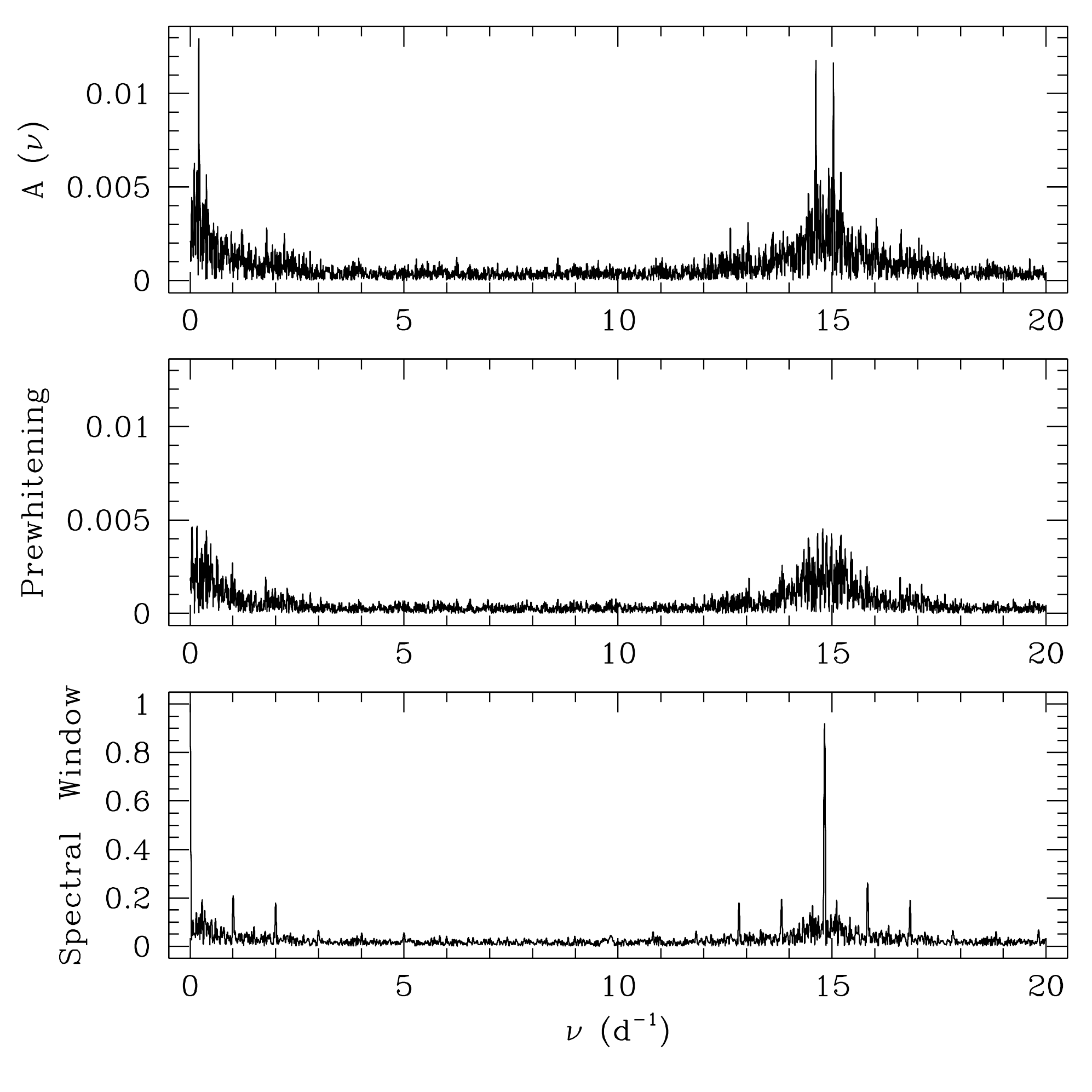}}
\end{center}
\end{minipage}
\hfill
\begin{minipage}{9cm}
\begin{center}
  \resizebox{9cm}{!}{\includegraphics{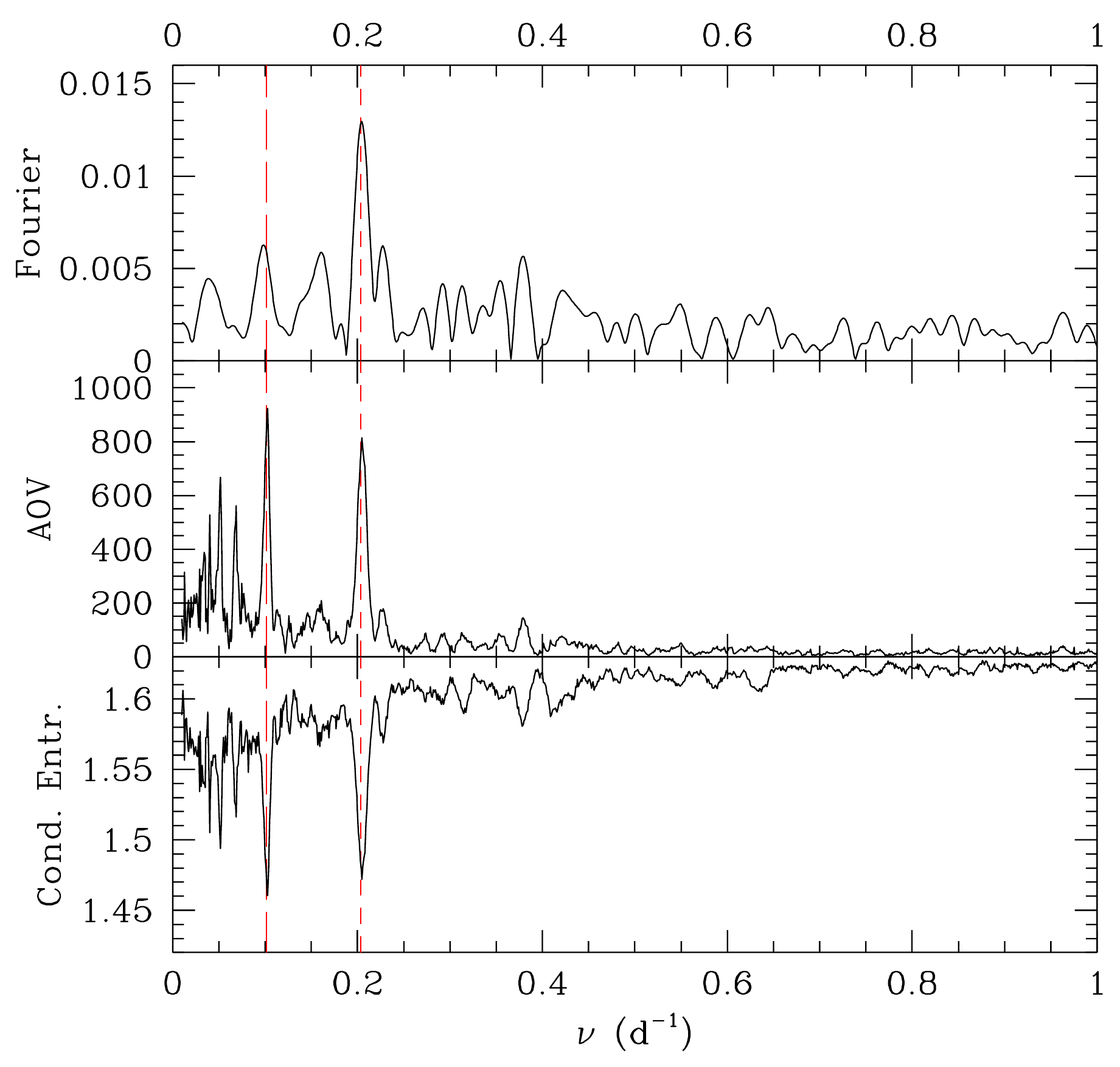}}
\end{center}
\end{minipage}
\caption{{\it Left}: Fourier periodogram of the decorrelated {\it BRITE} photometry of HD~149\,404 in the frequency range 0 to 20\,d$^{-1}$ (upper panel) and after pre-whitening of $P_{\rm orb}$ and $P_{\rm orb}/2$ (middle panel). Semi-amplitudes are expressed in magnitudes. The lower panel illustrates the spectral window. {\it Right}: Zoomed-in version of the frequency range between 0 and 1\,d$^{-1}$. The upper, middle, and lower panels illustrate the Fourier periodogram, the analysis of variance (AoV), and the conditional entropy methods, respectively. In the Fourier and AoV method a detection of a signal corresponds to a peak, whereas in the conditional entropy method a detection is indicated by a minimum. The long- and  short-dashed red lines correspond to $\nu_{\rm orb}$ and  $2 \times \nu_{\rm orb}$, respectively. \label{periodogram}}
\end{figure*}

The time series of SMEI data of HD~149\,404 is essentially continuous, except for a gap of about 2.5 months every year due to the location of the star near the ecliptic. The original SMEI photometry contains many outliers and is subject to a number of instrumental variations (see \citealt{Baade} for a description of these instrumental effects). The most prominent instrumental effect is a modulation by $\sim 0.5$\,mag with a 1-year period. To correct for this  effect, we first phase the raw data with a period of one year. Then, median values are calculated in 200 phase intervals. The annual effect is modelled via interpolation between these `anchor' points. We assume that the maximum of the interpolated curve indicates the correct flux of the source and add the missing flux to each data point using the interpolated 1-year lightcurve. To remove the outliers in the resulting lightcurve and account for the residual long-term instrumental trends, we apply the following iterative procedure:
\begin{itemize}
\item[$\bullet$] Mean magnitudes over time intervals of a duration $T$ are calculated. Following an initial detrending with $T = 100$\,day, we use values of $T$ which are multiples of the binary's orbital period, starting from five times the orbital period down to one orbital period. This strategy is adopted to avoid altering the orbital signal in the lightcurve. 
\item[$\bullet$] The mean magnitudes are interpolated and sigma clipping is  used to remove the outliers. We start with a large threshold (5\,$\sigma$) and make it smaller for the final iterations (3.5\,$\sigma$).
\item[$\bullet$] After a few iterations, the data were detrended for the binary orbital frequency and its harmonic, and the residuals of this fit were used for further sigma clipping.
\end{itemize}
The typical error bar on the final detrended SMEI photometry is 0.026\,mag.

\section{The HD~149\,404\label{lcBRITE} lightcurve}
The full detrended {\it BRITE} lightcurve of HD~149\,404 in the red filter is illustrated in the top panel of Fig.\,\ref{lcdetrend}. Variations with peak-to-peak amplitudes near 0.1\,mag are apparent.

\subsection{Frequency content of the lightcurve}
We analysed the decorrelated {\it BRITE-Heweliusz} time series of HD~149\,404 with a suite of period search algorithms including Fourier periodograms designed for uneven sampling \citep{HMM,Gosset}, trial period phase-diagram string length methods \citep{LK,Renson}, trial-period phase-diagram binned analyses of variances \citep{Whittaker,Jurkevich,Stellingwerf,Cuypers,SC}, and conditional entropy \citep{Cincotta,Cincotta2,Graham}. All these period search methods reveal strong signals at the binary's orbital period ($P_{\rm orb}$) and at its second harmonic ($P_{\rm orb}/2$). As an illustration, the Fourier periodogram, the analysis of variance and the conditional entropy are shown in the right panel of Fig.\,\ref{periodogram}. The entire Fourier power spectrum (top left panel of Fig.\,\ref{periodogram}) is clearly dominated by low frequencies ($< 3$\,d$^{-1}$). The orbital frequency of the {\it BRITE-Heweliusz} spacecraft (14.827\,d$^{-1}$) introduces a strong aliasing of these frequencies around 15\,d$^{-1}$ (see the spectral window function in the left panel of Fig.\,\ref{periodogram}). The Fourier periodogram is dominated by a progenitor peak at $2 \times \nu_{\rm orb}$ and a lower one at $\nu_{\rm orb}$. This is typical of the alternation of two minima of different depths. The power at the second harmonic frequency is mainly defined by the depth of the minima, whilst the power at the fundamental frequency mainly stems from the differences between the depths of the minima. In the analysis of variance (AoV) periodogram, the peak at the orbital frequency is more important because it is a blend of the genuine peak at $\nu_{\rm orb}$, which has a width equal to the natural width $\Delta\,\nu_{\rm nat}$, with a narrower subharmonic of $2 \times \nu_{\rm orb}$. The presence of peaks at subharmonic frequencies is a well-known artefact of phase-diagram methods. Several narrower subharmonic peaks are also visible around 0.05\,d$^{-1}$.

We pre-whitened the decorrelated data for $P_{\rm orb}$ and $P_{\rm orb}/2$ by fitting Eq.\,\ref{eqprew} to represent the orbital modulation. In this way, we find semi-amplitudes of $0.0067 \pm 0.0001$ and $0.0139 \pm 0.0001$\,mag for the variations at $P_{\rm orb}$ and $P_{\rm orb}/2$, respectively\footnote{ The errors on $A_j$ were determined from the interval of that parameter that leads to a variation of  $\frac{\chi^2\,n}{\chi^2_{\rm min}}$ by one unit, where $n$ is the number of degrees of freedom.}. 

The Fourier periodograms of the pre-whitened time series (see middle panel of Fig.\,\ref{periodogram}) indicate that any additional, non-orbital, variability in the photometric data occurs at frequencies below 3\,d$^{-1}$. There is no outstanding peak in the periodogram after pre-whitening the orbital variations, and for any given frequency the associated semi-amplitude is less than 5\,mmag. Moreover, the residual power decreases towards higher frequencies. This situation is reminiscent of the  red noise that was found in {\it CoRoT} photometry of several O-type stars: HD~46\,223 [O4\,V((f$^+$))], HD~46\,150 [O5.5\,V((f))], HD~46\,966 [O8\,V] \citep{Blomme}. Quite recently  a similar red noise component was also found in the {\it BRITE} photometry of the O4\,Ief supergiant $\zeta$~Pup \citep{Tahina}. We come back to this aspect in Sect.\,\ref{nonorb}. We note that this red noise is not of instrumental origin as it is only rarely seen in {\it BRITE} data. For instance, the periodogram of the {\it BRITE} data of the $\beta$~Cep pulsator $\sigma$~Sco, which was observed in the course of the same campaign as HD~149\,404, lacks any increase in the power towards lower frequencies \citep{sigSco}.\\

The absence of short-period variations implies that, for the study of the orbital variability, we can improve the signal-to-noise ratio of the data by averaging over the duration of the spacecraft orbit (0.06745\,day) without any loss of information. We confirmed this by checking that the periodograms of the data averaged over the spacecraft orbit remain essentially identical to those shown in Fig.\,\ref{periodogram}. The typical error bars on the data averaged over the spacecraft orbit is 0.002\,mag.

\begin{figure}[h]
\begin{center}
\resizebox{8cm}{!}{\includegraphics{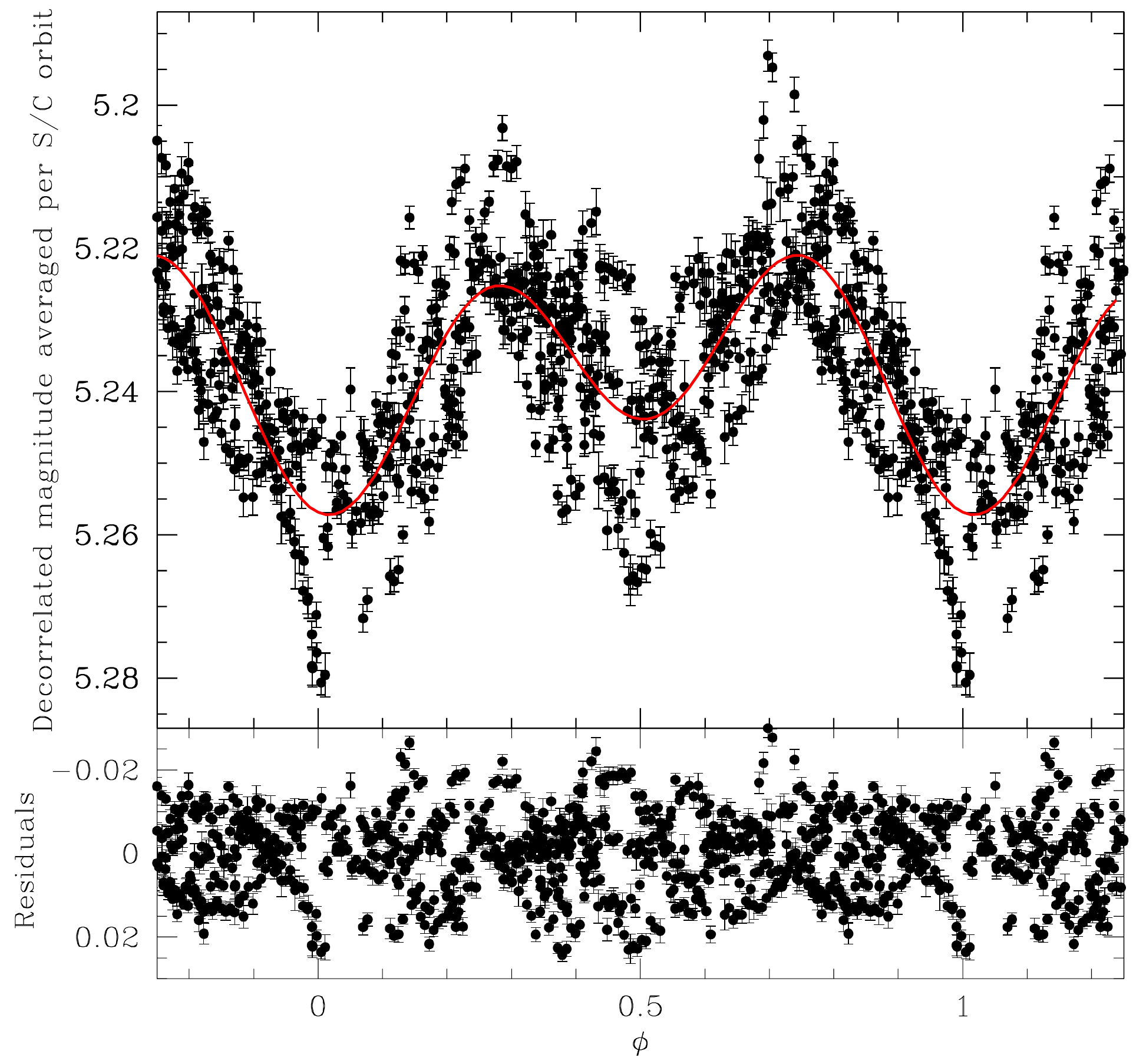}}
\end{center}  
\caption{Decorrelated {\it BRITE-Heweliusz} photometry of HD~149\,404, averaged over the spacecraft orbit, folded with the orbital period of the binary system. The red solid curve yields the best-fit Fourier reconstruction consisting of two sine waves: one at the orbital frequency and one at its second harmonic (see Eq.\,\ref{eqprew}). The lower panel illustrates the residuals between the data and this fit.\label{varorbit}}
\end{figure}

The lightcurve averaged over the duration of the spacecraft orbit contains a total of 708 data points (see the bottom panel of Fig.\,\ref{lcdetrend}). Figure\,\ref{varorbit} illustrates this lightcurve as a function of binary orbital phase computed according to the ephemeris of \citet{Rauw} where phase 0.0 corresponds to the ON9.7\,I secondary star being in inferior conjunction.  Ellipsoidal variations with a semi-amplitude near 0.02\,mag clearly dominate the lightcurve.

\begin{figure}[htb]
\begin{center}
  \resizebox{8cm}{!}{\includegraphics{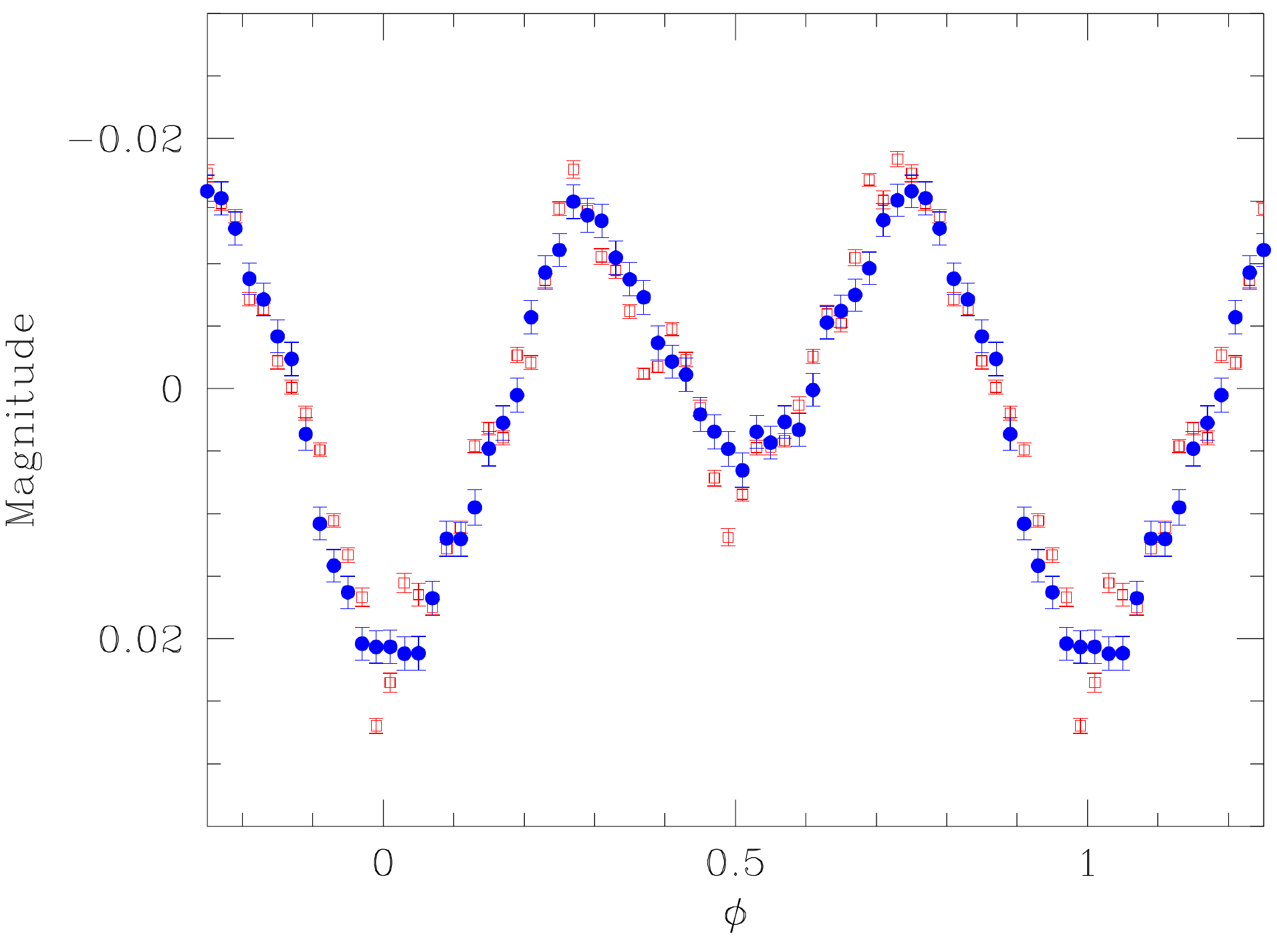}}
\end{center}  
\caption{Comparison between the {\it BRITE} (red open squares) and SMEI (filled blue dots) lightcurve folded with the orbital period of HD~149\,404. The data points correspond to normal points built adopting a phase bin of 0.02. To highlight the actual variations, the mean magnitudes are subtracted from both lightcurves.\label{lcnormal}}
\end{figure}
In the pre-whitening process (Eq.\,\ref{eqprew}), we fit the semi-amplitudes $A$ and phase constants $\psi$ of the sine waves corresponding to the orbital frequency and its second harmonic simultaneously for each of these frequencies. Applying the same method to the {\it BRITE} data averaged over the spacecraft orbit, we find that the orbital frequency and its second harmonic have semi-amplitudes of $A_1 = 0.0070 \pm 0.0005$\,mag and $A_2 = 0.0135 \pm 0.0005$\,mag, respectively. 
The resulting best-fit curve is illustrated in Fig.\,\ref{varorbit}. There are several important remarks to be made here:
\begin{itemize}
\item[$\bullet$] This best-fit curve obtained through Eq.\,\ref{eqprew} yields a difference between the depth of the minima near 0.0 and 0.5 of 0.013\,mag, and a difference in the maxima at phases 0.25 and 0.75 of 0.005\,mag (the maximum at phase 0.75 being brighter). This small difference between the maxima is not significant in view of the non-orbital variability that affects the lightcurve (see Sect.\,\ref{nonorb}).\\
  
\item[$\bullet$] According to this best fit, there is a slight shift of the phase of primary minimum compared to the ephemeris of \citet{Rauw} of 0.02, equivalent to 0.196\,days. This shift could result from the uncertainty of $P_{\rm orb}$ accumulated over the long time interval (5525 days, i.e.\ 563 orbital cycles) between the spectroscopic data of \citet{Rauw} and the photometric data analysed here, or from the uncertainty of the time of conjunction $T_0$ given by \citet[][see Table\,\ref{param}]{Rauw}, or from a combination of both effects. Since we cannot disentangle these possible causes of the shift, we simply applied a phase shift of 0.02 in our analysis of the lightcurve (Sect.\,\ref{modelorb}). This leads to a revised value of $T_0 = 2\,457\,215.9943$. An alternative would be a somewhat increased orbital period (9.81510\,days, i.e.\ an adjustment of 0.0036\% of $P_{\rm orb}$).
\end{itemize}

\begin{figure*}
\begin{center}
  \resizebox{16cm}{!}{\includegraphics{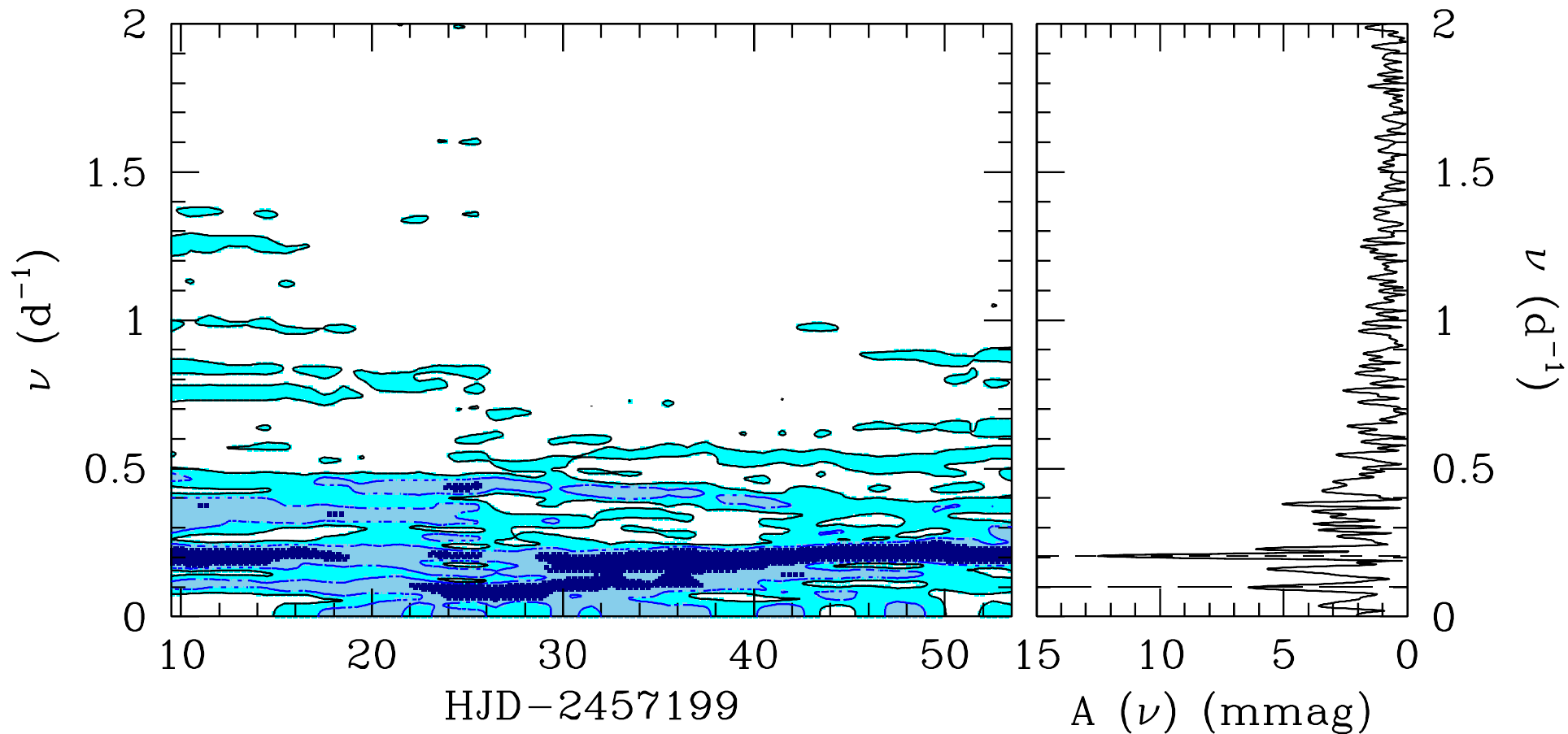}}
  \resizebox{16cm}{!}{\includegraphics{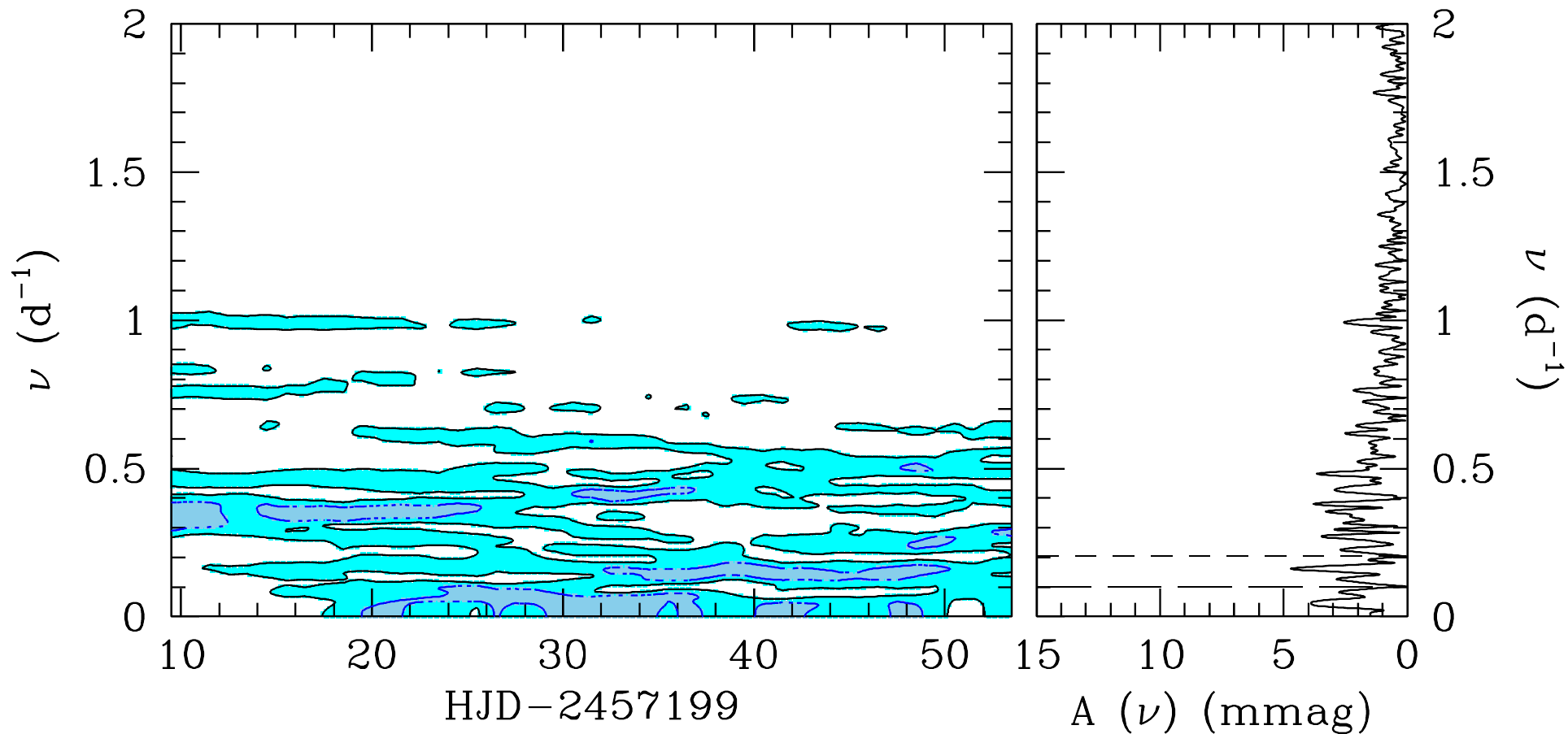}}
\end{center}  
\caption{Time-frequency diagrams of the decorrelated photometry of HD~149\,404, averaged over the spacecraft orbit before (upper panels) and after (lower panels) pre-whitening for $P_{\rm orb}$ and $P_{\rm orb}/2$. The left panels illustrate the evolution of the Fourier periodogram as a function of the heliocentric Julian day at the middle of the 20-day sliding window (see text). Dark blue, light blue, and cyan stand respectively for semi-amplitudes $\geq 9$\,mmag, $\geq 6$\,mmag, and $\geq 3$\,mmag. The right panels yield the Fourier periodograms over the full duration of the {\it BRITE} campaign.\label{spevol}}
\end{figure*}
In a binary system with a circular orbit, such as HD~149\,404, a difference in depth of the photometric minima could result from several factors. A first possibility would be eclipses of stars of unequal surface temperature. As an alternative, mutual reflection effects between distorted stars of unequal surface brightness will result in ellipsoidal variations with a deeper minimum when the hotter star is in superior conjunction. Finally, the properties of the orbital variations could hint at an asymmetry in the system. Such an asymmetry could result from a `spot' on the surface of one of the stars \citep[e.g. a bright spot could stem from the impact of the primary stellar wind on the secondary star; see][]{Rauw,Naze,Thaller}.

However, before we accept these asymmetries as true, it is important to understand whether these deviations from a purely symmetrical ellipsoidal modulation are real or result from interference with the remaining frequency content of the lightcurve. For this purpose, we turn to the data collected with the SMEI instrument. Although the SMEI data have much larger error bars than the {\it BRITE} data and are subject to possible contamination by neighbouring sources, their large number (21\,906 data points after removing the outliers) and their very long time span offer an interesting complement to the {\it BRITE} data. To illustrate this point, we compare in Fig.\,\ref{lcnormal} the {\it BRITE} and SMEI lightcurves phase-folded over the orbital cycle of HD~149\,404 and consisting of normal points built by adopting a phase bin of 0.02. Both curves are in very good agreement. The SMEI lightcurve displays slightly lower dispersion, probably because the SMEI data cover more than 200 orbital cycles of HD~149\,404 making the normal points less sensitive to the effect of non-orbital variations (see Sect.\,\ref{nonorb}).

Appendix\,\ref{appendix1} provides detailed information on the periodograms built from the SMEI data. The orbital frequency and the second harmonic clearly dominate the periodogram (Figs.\,\ref{periodogramSMEI1} and \ref{periodogramSMEI2}). For these data, we determine semi-amplitudes of the modulations of $A_1 = 0.0078 \pm 0.0002$\,mag and $A_2 = 0.0135 \pm 0.0002$\,mag for the orbital period and its second harmonic, respectively. The best-fit two-frequency relation for the SMEI data essentially confirms the conclusions drawn from the {\it BRITE} lightcurve with regard to the depth of the minima. Indeed, for the SMEI data, the difference between the depths of the minima at $\phi = 0.0$ and $\phi = 0.5$ amounts to $\sim 0.015$\,mag. However, the asymmetry between the maxima at quadrature phases appears  less pronounced in the SMEI data where the maximum at $\phi = 0.75$ is only $\sim 0.001$\,mag brighter than that at $\phi = 0.25$.   

To get further insight into the origin of the slight differences between the {\it BRITE} and SMEI lightcurves, we performed a Fourier analysis of the {\it BRITE} photometric time series, averaged over the duration of the spacecraft orbit, cut into sliding windows of twenty days. The length of the sliding window was chosen to cover at least two full orbital cycles. The resulting time-frequency diagrams are shown in Fig.\,\ref{spevol}. 

From Fig.\,\ref{spevol} it appears that the frequency corresponding to $P_{\rm orb}/2$ is always present, though with an apparently lower semi-amplitude between days 19 and 28. The frequency corresponding to $P_{\rm orb}$ has a more irregular visibility: whilst it is present at the $\geq 6$\,mmag level most of the time, it gets stronger than 9\,mmag between days 22 and 37 (partially overlapping with the episode of reduced strength of the $P_{\rm orb}/2$ signal) and drops below 6\,mmag after day 41. It seems thus that both frequencies are present most of the time, but their relevance seems to be affected by the non-orbital variations. 

\subsection{Non-orbital variability \label{nonorb}}
Within the level of accuracy of the {\it BRITE} data, the periodograms of the HD~149\,404 photometry reveal no obvious discrete peaks at frequencies other than those corresponding to the orbital modulation. In particular, we note that there are no signals in the range of periods typically associated with $\beta$~Cep-like pulsations \citep[which span a range from $\sim 0.1$\,day to $\sim 0.25$\,day with a median value near $0.17$\,day,][]{PP}. Instead, as pointed out above, the frequency content of the non-orbital variations of HD~149\,404 is essentially consistent with red noise. In the case of early-type stars, this designation is used for apparently stochastic signals whose power increases towards lower frequencies. Several scenarios have been proposed to explain the origin of these variations. They include unresolved randomly excited p-mode oscillations or internal gravity waves generated either in a subsurface convection zone or directly in the convective core of a massive star \citep[see][and references therein]{Tahina}.

Though the power of the red noise is spread out over a range of frequencies, the residuals seen in Fig.\,\ref{varorbit} indicate that it can produce variations with peak-to-peak amplitudes up to 40\,mmag. This is larger than the 20\,mmag peak-to-peak amplitude reported by \citet{Tahina} in the case of $\zeta$~Pup.

\begin{figure}[h]
\begin{center}
\resizebox{8cm}{!}{\includegraphics{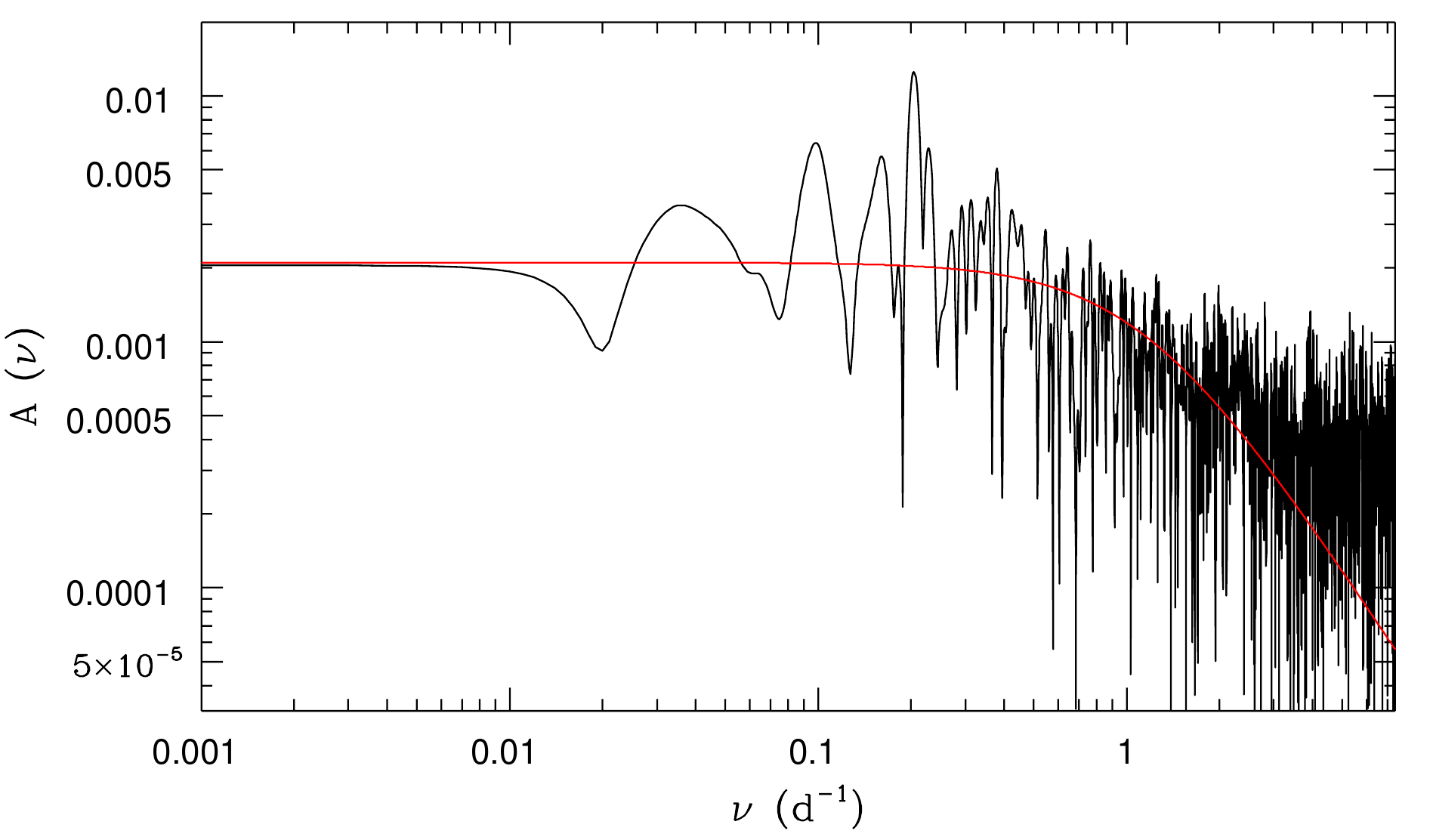}}
\end{center}  
\caption{Log-log plot of the semi-amplitude spectrum of the decorrelated photometry, averaged over the spacecraft orbit. The semi-amplitude spectrum is shown out to the Nyquist frequency of the spacecraft orbit (7.414\,d$^{-1}$). The red line corresponds to the best-fit red-noise relation with $A_0 = 0.0021$, $\gamma = 1.93 \pm 0.11$ and $\tau = 0.138 \pm 0.005$\,day, fit up to a frequency of 4.0\,d$^{-1}$.}
\label{rnfit}
\end{figure}

We model the non-orbital variations following the approach of \citet{Blomme} and \citet{Uuh}, which is based on the formalism of \citet{Stanishev}. We thus fit the expression
\begin{equation}
  A(\nu) = \frac{A_0}{1 + (2\,\pi\,\tau\,\nu)^{\gamma}}
  \label{eq1}
\end{equation}
to the periodogram (see Fig.\,\ref{rnfit}). Here $A(\nu)$ is the semi-amplitude derived from the Fourier periodogram, $A_0$ is a scaling factor, $\gamma$ is the slope of the linear part (in a log-log plot), and $\tau$ (in days) is an indication of the mean duration of the dominant features in the lightcurve. We perform the fit up to a frequency of 4.0\,d$^{-1}$. At higher frequencies, up to 7.414\,d$^{-1}$ (i.e.\ up to the Nyquist frequency of the spacecraft orbit), the periodogram is essentially consistent with white noise.

\begin{figure}[h]
\begin{center}
\resizebox{8cm}{!}{\includegraphics{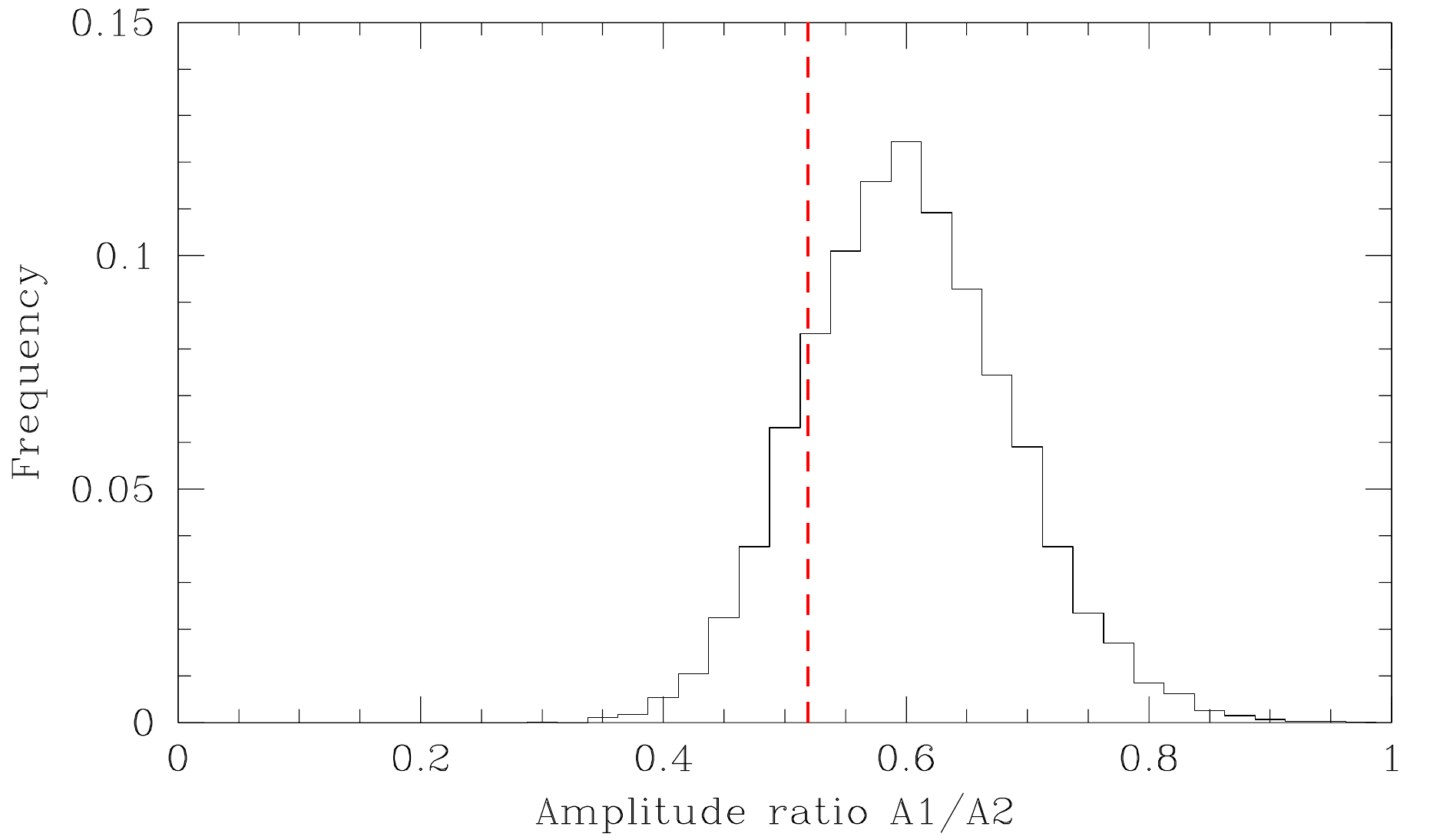}}
\end{center}
\begin{center}
\resizebox{8cm}{!}{\includegraphics{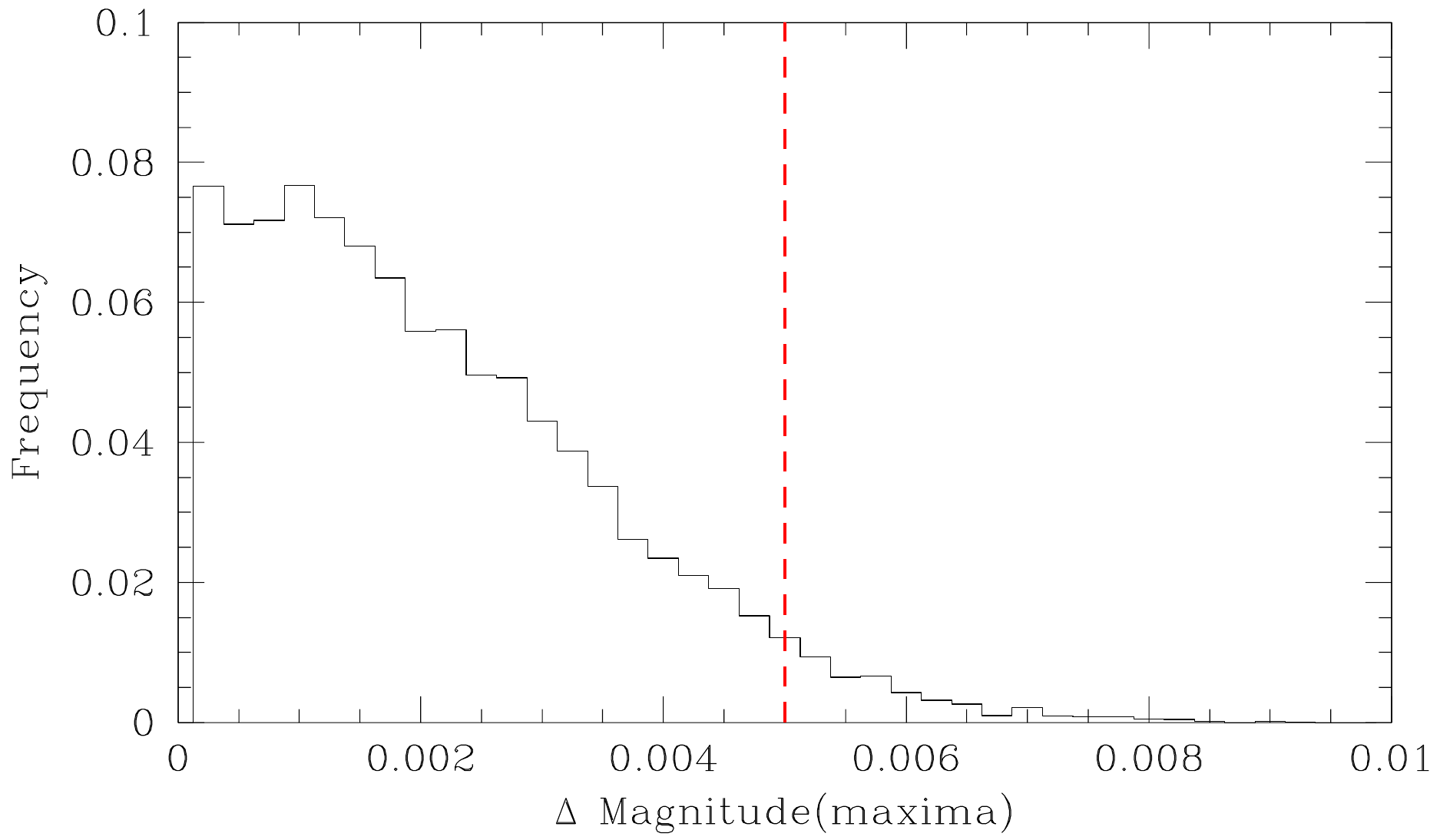}}
\end{center}
\caption{Top: Histogram of the ratio between the semi-amplitudes of the orbital frequency and its second harmonic as obtained with the Monte Carlo simulations. The dashed vertical line indicates the observed value of the semi-amplitude ratio obtained via pre-whitening of the lightcurve averaged over the {\it BRITE} spacecraft orbit $(A_1/A_2)_{\rm BRITE} = 0.52 \pm 0.05$. Bottom: Histogram of the difference between the two maxima of the simulated lightcurves. The observed value of this difference in the {\it BRITE} lightcurve ($\simeq 0.005$\,mag) is again shown by the dashed vertical line.\label{histogram}}
\end{figure}

Our best-fit $\gamma$ value ($1.93 \pm 0.11$) is  between the values (0.91 -- 0.96) of the three early-type stars investigated by \citet{Blomme} and the value (2.3) obtained for Plaskett's Star by \citet{Mahy}. The $\tau$ value ($0.138 \pm 0.005$\,day) also falls  in the range observed for these stars \citep[0.08 -- 0.17\,day,][]{Blomme,Mahy}.

To assess the impact of the intrinsic variations on the orbital modulation, and more specifically to understand whether these intrinsic variations can explain the slight differences between the {\it BRITE} and SMEI lightcurves, we generated 10\,000 synthetic {\it BRITE} lightcurves using a Monte Carlo simulator. We assume that the orbital signal consists of the best-fit two frequency decomposition of the SMEI lightcurve and a red noise component simulated using the recipe of \citet{TK} with the $A(\nu)$ formalism described in Eq.\,\ref{eq1} above. We adopt the same sampling as for the spacecraft-orbit averaged {\it BRITE} lightcurve. We then fit an expression equivalent to Eq.\,\ref{eqprew} to each synthetic lightcurve. The results are illustrated by the two histograms in Fig.\,\ref{histogram}. The top panel displays the ratio between the semi-amplitudes of the sine waves corresponding to the orbital period ($A_1$) and its second harmonic ($A_2$). As can be seen, the distribution of the $A_1$-to-$A_2$ ratios for the simulated data peaks near the value determined for the SMEI data, $(A_1/A_2)_{\rm SMEI} = 0.58$, which we used as input in our simulations. The value determined from the {\it BRITE} data, $(A_1/A_2)_{\rm BRITE} = 0.52 \pm 0.05$, is very close to the peak of the distribution of the simulated data and is thus not exceptional given the properties of the red noise.

The bottom panel of Fig.\,\ref{histogram} illustrates the difference in magnitude between the two maxima of the simulated lightcurves. Among the 10\,000 synthetic lightcurves, 568 yield a difference between the maxima that is equal to or larger than the observed value. We thus conclude that, in view of the non-orbital variations, the significance level of the observed difference of the maxima is about 5.7\%. The slight asymmetry of the {\it BRITE} lightcurve around the maxima is thus likely insignificant.

From these simulations, we  conclude that the red noise component can affect the apparent asymmetry of the orbital component of the lightcurve of HD~149\,404 and can thus explain the slight differences between the fits of the {\it BRITE} and SMEI lightcurves. In Sect.\,\ref{modelorb}, we  consider various options to extract and model the orbital part of the lightcurve.   
 
\section{Modelling the orbital variations\label{modelorb}}
In view of the above discussion about possible asymmetries in the orbital lightcurve, we considered several approaches to fit the orbital part of the photometric variations of HD~149\,404. We first focused on the decorrelated {\it BRITE} photometry averaged over the spacecraft orbit (option I). Then we analysed the lightcurves consisting of normal points built by adopting phase bins of 0.02 (option II and III for the {\it BRITE} and SMEI lightcurves, respectively). 

We analysed the lightcurve with the eclipsing binary star code {\tt nightfall} (version 1.86) developed by R.\ Wichmann, M.\ Kuster and P.\ Risse\footnote{The code is available at the URL: http://www.hs.uni-hamburg.de/DE/Ins/Per/Wichmann/Nightfall.html}. This code uses the Roche potential to describe the shape of the stars, and for the simplest cases (two stars on a circular orbit with no stellar spots or disks), six parameters are required to describe the system:  the mass-ratio, the orbital inclination, the primary and secondary filling factors (defined as the ratio of the stellar polar radius to the polar radius of the associated Roche lobe), and the primary and secondary effective temperatures. We fixed the mass-ratio and the stellar effective temperatures to the values obtained in the spectroscopic analyses of \citet{Rauw} and \citet{Raucq} listed in Table\,\ref{param}. We further adopted a quadratic limb-darkening law using limb-darkening coefficients from \citet{Claret}. Reflection effects were accounted for by considering the mutual irradiation of all pairs of surface elements of the two stars \citep{Hendry}. To approximate the non-standard passbands of the {\it BRITE} and SMEI instruments in the {\tt nightfall} calculations, we used the Johnson-$V$ passband as a proxy. Whilst this is clearly an approximation, it should not impact our conclusions as both components of HD~149\,404 are relatively close in spectral type and thus have spectral energy distributions in the optical that are quite similar. Finally, since we are analysing broadband photometry in a single passband, we used the black-body law to compute the relative contribution from each star to the total light. 

The spectroscopic optical brightness ratio of the system (see Table\,\ref{param}) can be converted into a bolometric luminosity ratio of $L_{\rm bol,1}/L_{\rm bol,2} = 1.11 \pm 0.19$ adopting the bolometric corrections inferred by \citet{Raucq}. Then, knowing the effective temperatures, we obtain the ratio between the stellar radii ($R_1/R_2 = 0.713 \pm 0.116$) and hence between the Roche lobe filling factors: $f_1/f_2 = 0.566 \pm 0.093$.

\begin{figure*}[h!tb]
  \begin{minipage}{8cm}  
    \resizebox{8cm}{!}{\includegraphics{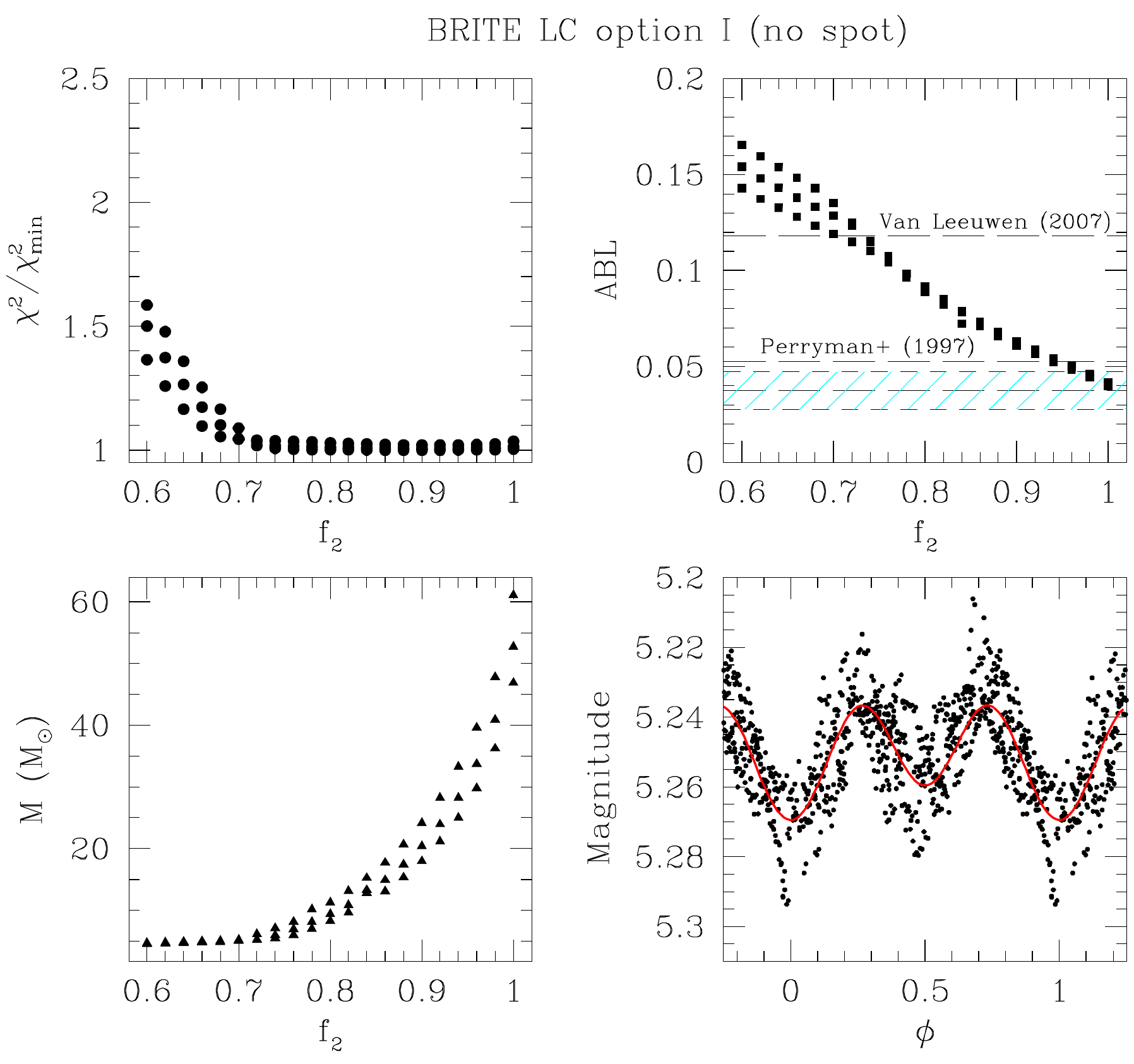}}
  \end{minipage}
  \hfill
  \begin{minipage}{8cm}
    \resizebox{8cm}{!}{\includegraphics{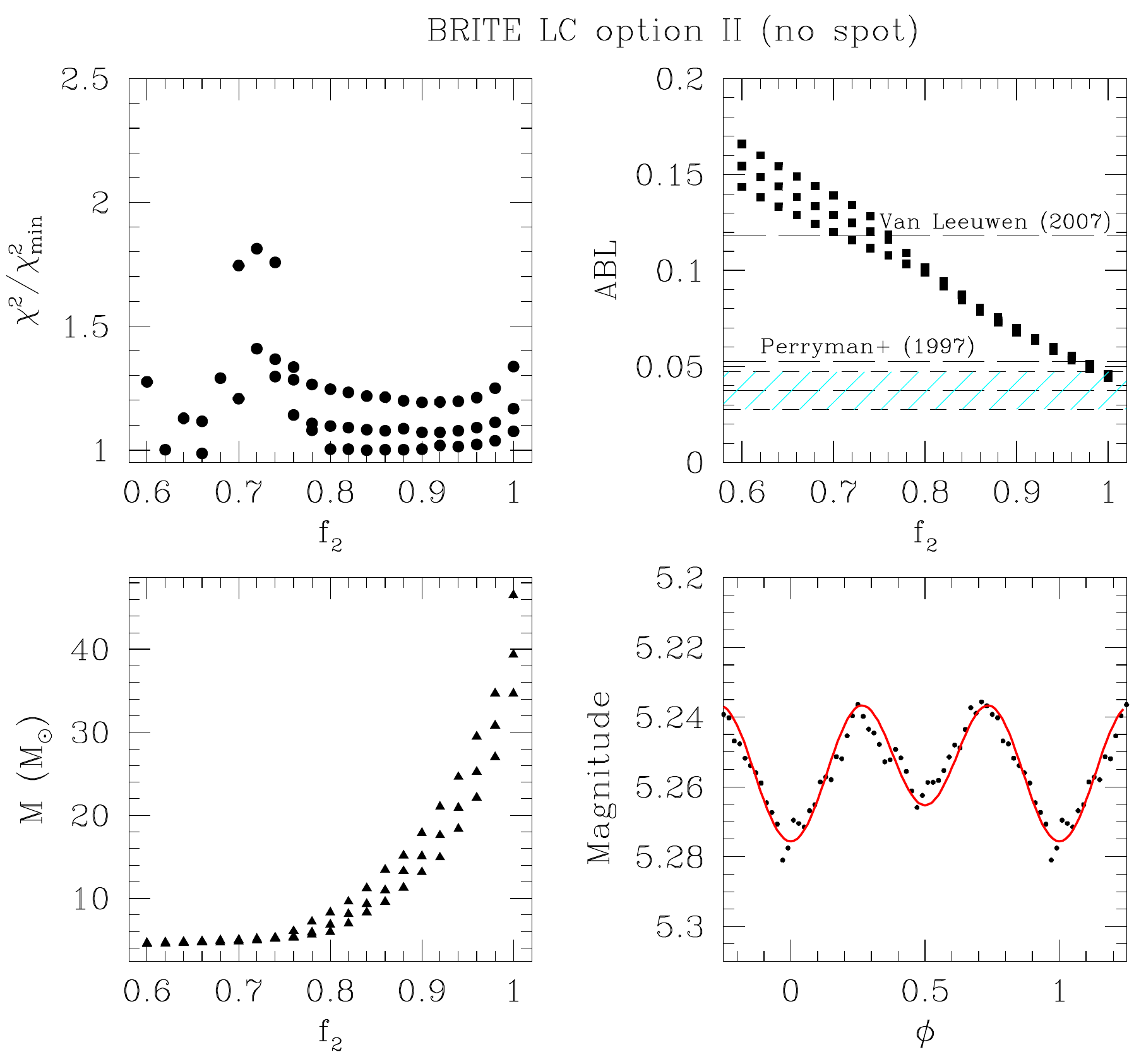}}
  \end{minipage}
  \hfill
  \begin{minipage}{8cm}
    \resizebox{8cm}{!}{\includegraphics{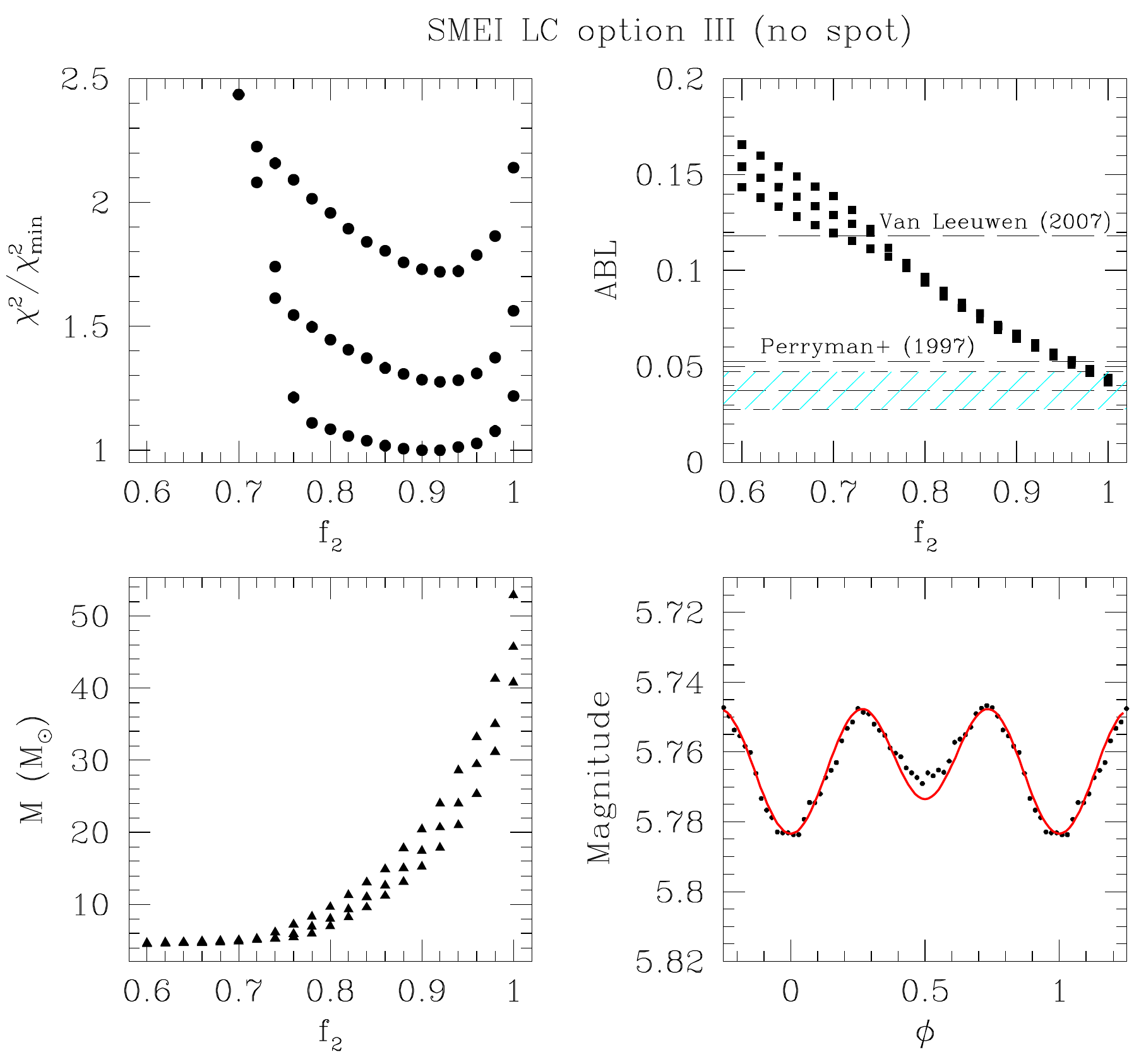}}
      \end{minipage}
  \hfill
  \begin{minipage}{8cm}
    \caption{Properties of the best-fit models (assuming no surface spots) of the lightcurves for option I (spacecraft-orbit-averaged {\it BRITE} photometry, upper left quadrant), option II (binned {\it BRITE} lightcurve, upper right quadrant), and option III (binned SMEI lightcurve, lower left quadrant).
For each option, the panels illustrate the fit quality, the astrometry-based luminosity $ABL$ (see text), and the total mass of the system as a function of the secondary filling factor. The blue hatched area in the $ABL$ panels indicates the value of $ABL_{\rm DR2}$ and its associated uncertainty. The lower right panel of each quadrant compares the lightcurve with the fit with the lowest $\chi^2$ that also agrees with the value of $ABL_{\rm DR2}$ (see Table\,\ref{table1}).\label{propsols}}
  \end{minipage}
\end{figure*}

\subsection{Models without spots \label{wospot}}
As a first step, we left the primary and secondary Roche lobe filling factors and the orbital inclination unconstrained in the fitting procedure. The best fits obtained in this way yield primary filling factors that are close to, or even larger than, the secondary filling factor. These solutions are thus significantly at odds with the ratio of the filling factors inferred from the spectroscopic analysis ($f_1/f_2 = 0.566 \pm 0.093$, see above). 

To search for solutions that are consistent with the spectroscopic ratio of stellar radii, we thus constrained the values of the filling factors according to the considerations above. For this purpose, we scanned the parameter space in a systematic way. For each value of $f_2$ between 0.60 and 1.00 (in steps of 0.02), we tested three values of $f_1$, corresponding respectively to ratios $f_1/f_2 - \sigma_{f_1/f_2}$, $f_1/f_2$, and $f_1/f_2 + \sigma_{f_1/f_2}$.
With these combinations of filling factors fixed in the fitting procedure, the only remaining fitting parameter is the orbital inclination. We thus ran the {\tt nightfall} code to search for the orbital inclination that yields the best-quality fit. As a first step, we included the detailed treatment of the mutual reflection between the stars, but omitted any spots on the stars. 

The results are illustrated in Fig.\,\ref{propsols} for the three options of the lightcurve defined above. For each option, Fig.\,\ref{propsols} provides the $\chi^2/\chi^2_{\rm min}$, the astrometry-based luminosity \citep{AL}
\begin{equation}
  ABL = 10^{0.2\,M_V}
  \label{ABL1}
,\end{equation}
where $M_V$ is the absolute magnitude of the system inferred from the lightcurve solution, the total mass of the binary, and a comparison between the best-fit solution that has an $ABL$ in agreement with the {\it Gaia}-DR2 parallax (see below) and the lightcurve.

\begin{figure}[h]
\begin{center}
\resizebox{8cm}{!}{\includegraphics{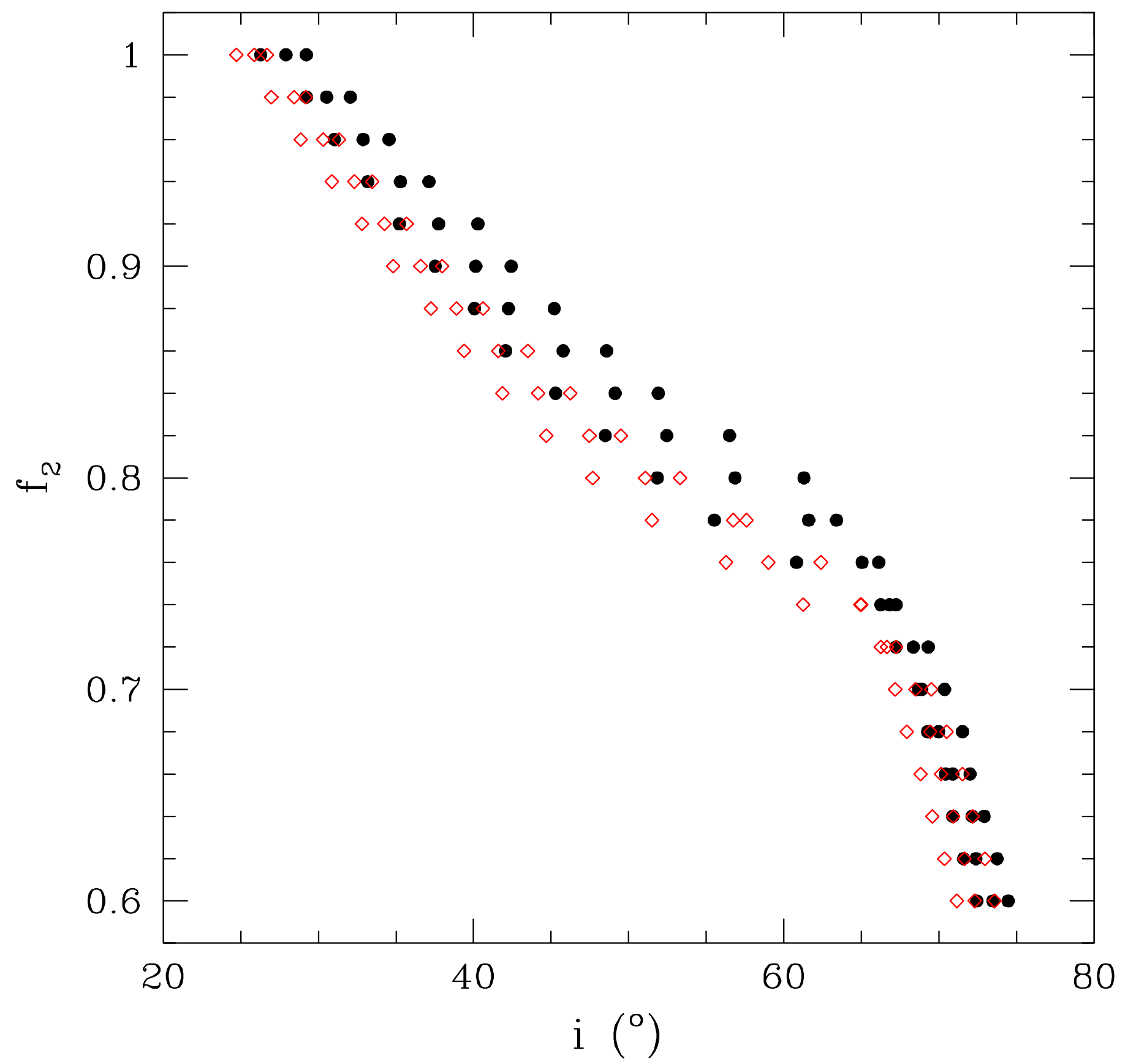}}
\end{center}  
\caption{Secondary filling factor $f_2$ as a function of orbital inclination for the best-fit lightcurve models of option II (binned {\it BRITE} lightcurve) without a spot on the secondary (filled dots) and including  a spot (red open diamonds). For each value of $f_2$ three models are shown, corresponding to the three values of $f_1$ investigated.}
\label{inclvsf2}
\end{figure}

There are several results that are worth noting here. First of all, the larger the secondary filling factor, the lower the orbital inclination required to reproduce the observed lightcurves (see Fig.\,\ref{inclvsf2}). For $f_2 = 1.00$, we find the lowest values of $i$ which range between $23$ and $26^{\circ}$, depending on the option of the lightcurve. Values of $f_2 \leq 0.70$ are generally associated with $i \geq 67^{\circ}$ and the corresponding synthetic lightcurves display grazing eclipses or even eclipses for the lowest values of $f_2$. Except for the {\it BRITE} lightcurve built from the normal points (option II), the presence of these (grazing) eclipses leads to a rapid degradation of the fit quality. In general, the values of the $\chi^2$ of the fits are far from what is expected for a statistically good fit\footnote{These values are nevertheless very close to those found for the best solutions with $f_1$ and $f_2$ left unconstrained. Hence, constraining the ratio $f_1/f_2$ does not impact the fit quality too much.}. For lightcurve option I and probably also option II, this situation reflects the presence of the strong red noise component on top of the ellipsoidal variations (see Sect.\,\ref{nonorb}). 
In Sect.\,\ref{wspot}, we investigate how the fit quality changes when we include a spot on the secondary surface. Meanwhile, we note that it is generally difficult to identify an overall best-fit solution of the lightcurve based on the behaviour of the $\chi^2$ value alone.  

Additional constraints can be obtained from the dependence of the absolute magnitude on $f_2$. Indeed, once we have obtained a full solution of the lightcurve, we can combine the Roche lobe filling factors with the orbital solution of \citet{Rauw} and the bolometric corrections and effective temperatures of \citet{Raucq} to compute the absolute magnitude of the HD~149\,404 binary system (see the parameters in Table\,\ref{param}). This absolute magnitude strongly varies as a function of $f_2$, and can thus be compared to the absolute magnitude based on the apparent magnitude and a distance estimate. For this purpose, we use the apparent magnitude $m_V = 5.48 \pm 0.02$ and colour-index $B-V = 0.39 \pm 0.02$ \citep{Reed}. We further adopt the intrinsic colour $(B-V)_0 = -0.26$ \citep{MP} and assume $R_V = 3.1$. Using the {\it Hipparcos} data, \citet{Perryman} quoted a parallax of $\varpi = 1.07 \pm 0.89$\,milliarcsec, whilst the revised reduction of the {\it Hipparcos} data led to a parallax of $\varpi = 2.40 \pm 0.36$\,milliarcsec \citep{VanLeeuwen}. A significantly larger distance ($1753 \pm 293$\,pc, corresponding to $\varpi = 0.57 \pm 0.10$\,milliarcsec) was estimated by \citet{Megier} based on the strength of the interstellar Ca\,{\sc ii} H and K lines. Most recently, the second {\it Gaia} data release \citep[DR2,][]{Brown} quoted a parallax of $\varpi = 0.760 \pm 0.165$\,milliarcsec. This value overlaps within the errors with the original {\it Hipparcos} parallax and the estimate of \citet{Megier}, but rules out the value from the revised {\it Hipparcos} reduction. Whilst the DR2 certainly provides the most accurate determination of the parallax of HD~149\,404 to date, we need to be careful when comparing our results with the DR2.  \citet{Luri} pointed out that a Gaussian probability density function (PDF) of the measured parallax $\varpi$ results in a non-Gaussian PDF for the distance estimator $d = 1/\varpi$. Even worse, the shape of the PDF of $d$ depends on the unknown value of the true parallax $\varpi_{\rm true}$. As a way to avoid this problem, \citet{Luri} suggest  expressing the quantity to be measured directly in the space of parallaxes. In our case, this can be done using the astrometry-based luminosity $ABL$ \citep{AL}. According to Eq.\,\ref{ABL1}, it follows that
\begin{equation}
  ABL = 10^{0.2\,M_V} = \varpi\,10^{0.2\,(m_V-A_V+5)}
\end{equation}
For HD~149\,404, the DR2 parallax results in $ABL_{\rm DR2} = (3.74 \pm 0.98)\,10^{-2}$. This value and its associated error are shown by the hatched area in the plots of Fig.\,\ref{propsols}. In a recent study, \citet{ST} compared the {\it Gaia}-DR2 parallaxes of 89 eclipsing binaries against the distances derived from analyses of the lightcurves. \citet{ST} found evidence for a systematic offset by $0.082 \pm 0.033$\,milliarcsec with the {\it Gaia}-DR2 parallaxes being too small. Accounting for such a systematic shift, does not affect our conclusion regarding the value of $f_2$: only solutions with $f_2 \geq 0.96$ agree with the $ABL_{\rm DR2}$ value.

\begin{table}[h]
  \caption{Parameters of the best fit with a simple binary model consistent with $ABL_{\rm DR2}$\label{table1}}
  \begin{center}
  \begin{tabular}{c c c c c}
    \hline
    LC option & $f_1$ & $f_2$ & $i$\,($^{\circ}$) & $\chi^2$ \\
    \hline
    I   & 0.646 & 0.980 & 28.8 & 28.6\\
    II  & 0.659 & 1.000 & 29.2 & 58.3\\
    III & 0.646 & 0.980 & 30.4 &  4.4\\ 
    \hline
  \end{tabular}
  \tablefoot{The models account for mutual reflection, but do not include any spots. The corresponding fits are illustrated in Fig.\,\ref{propsols}. The last column quotes the normalized $\chi^2$ of the fit. For comparison, the $\chi^2_{\rm min}$ values for the full parameter space that we explored are 28.6, 54.2, and 4.0 for lightcurve options I, II, and III, respectively.}
  \end{center}
\end{table}

\begin{figure*}[h!tb]
  \begin{minipage}{8cm}  
    \resizebox{8cm}{!}{\includegraphics{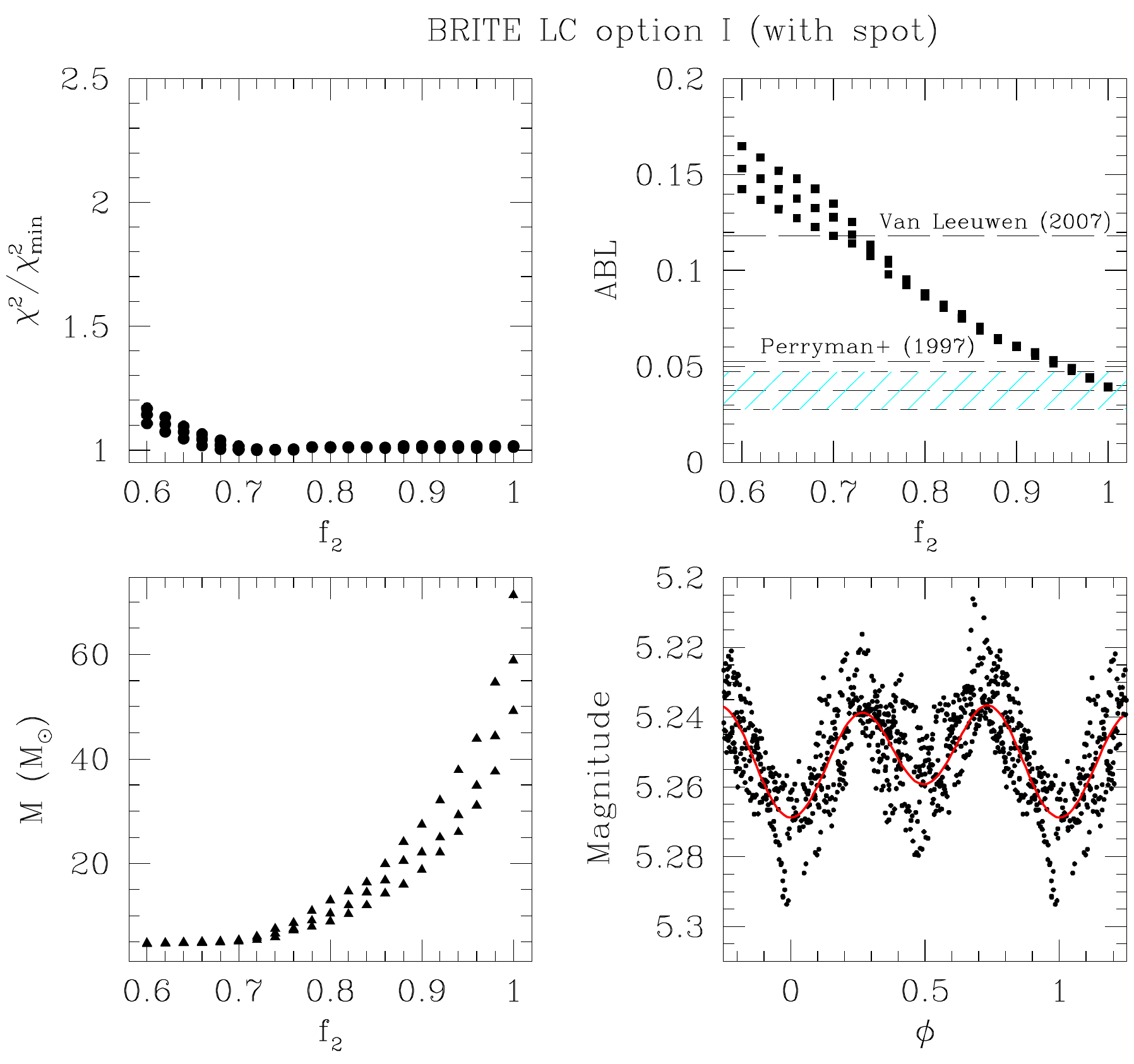}}
  \end{minipage}
  \hfill
  \begin{minipage}{8cm}
    \resizebox{8cm}{!}{\includegraphics{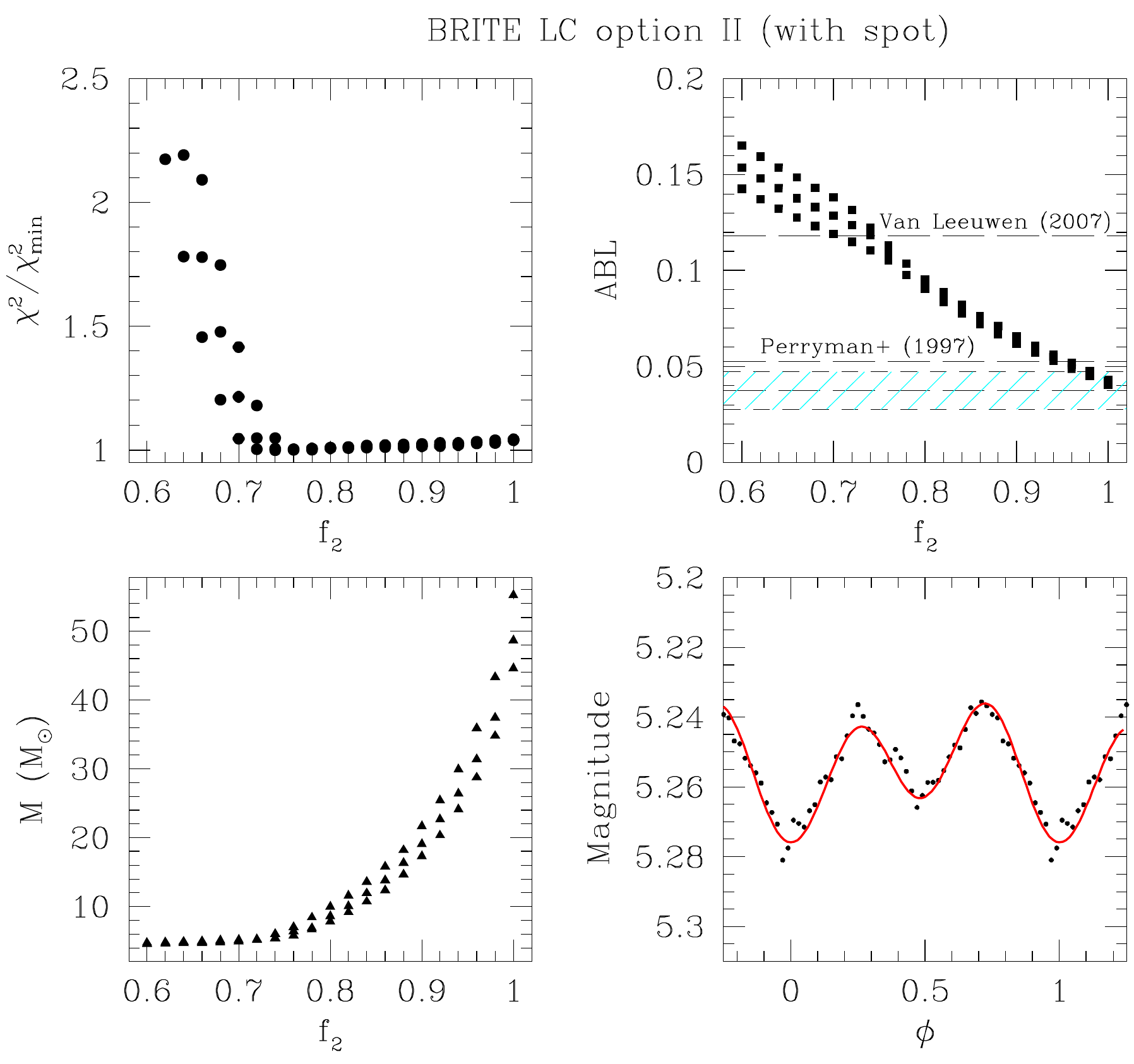}}
  \end{minipage}
  \hfill
  \begin{minipage}{8cm}
    \resizebox{8cm}{!}{\includegraphics{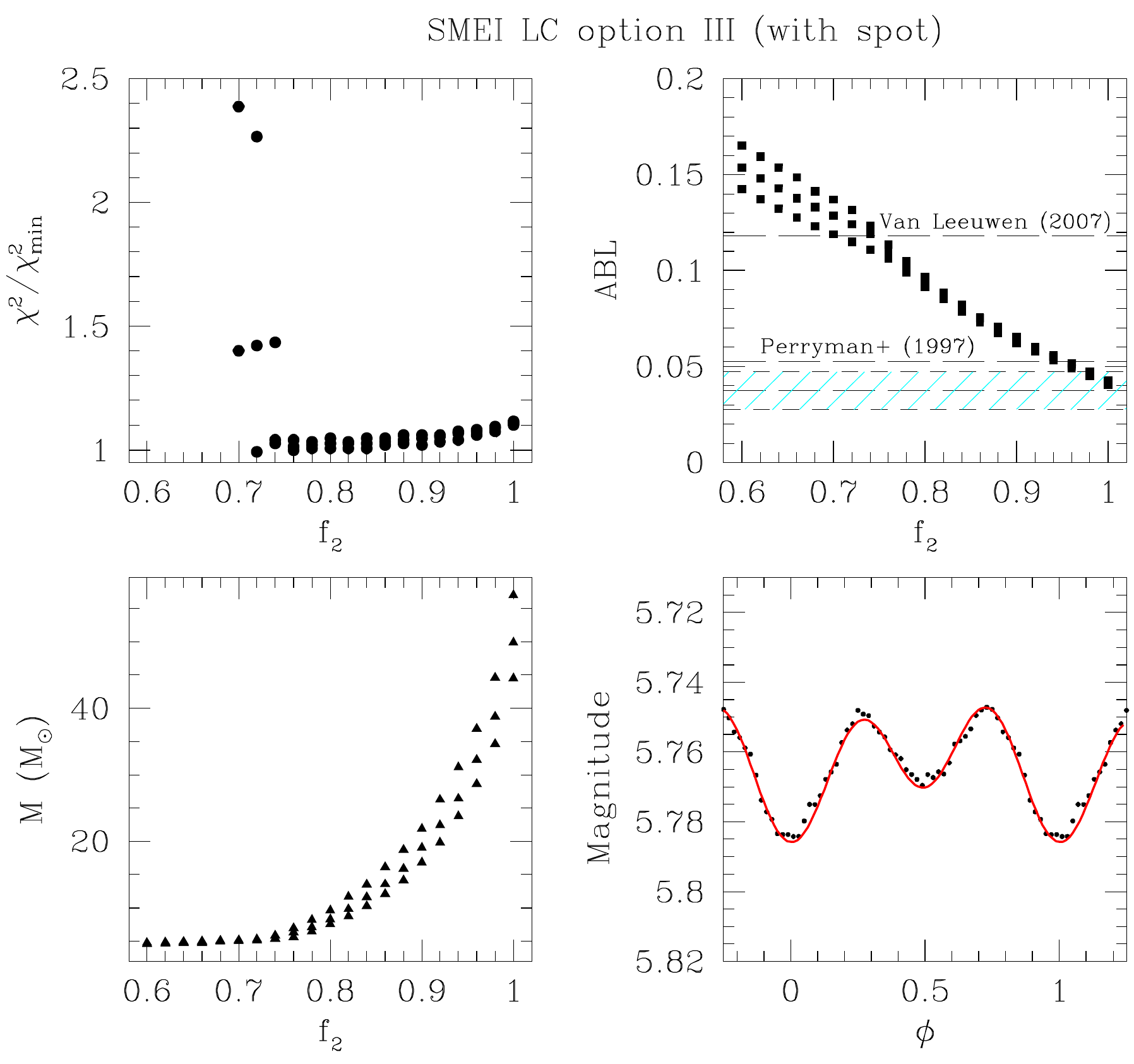}}
      \end{minipage}
  \hfill
  \begin{minipage}{8cm}
    \caption{Same as Fig.\,\ref{propsols}, but for the fits including a spot on the secondary surface. \label{propsols2}}
  \end{minipage}
\end{figure*}

\subsection{Models with spots \label{wspot}}
We also test models with a spot on the surface of the secondary. We chose to put the spot on the secondary star because the analyses of the emission line profile variations of \citet{Rauw}, \citet{Naze}, and \citet{Thaller} suggest that the wind of the primary dominates over that of the secondary, possibly leading to a crash of the primary wind onto the surface of the secondary. The spot is assumed to be located on the secondary's equator, but we leave its size and longitude as free parameters. The spot can be either brighter or fainter than the surrounding stellar surface. In this way, we now have for each combination of $f_1$ and $f_2$, a total of four free parameters: the orbital inclination $i$, plus the longitude, radius, and dim-factor\footnote{The dim-factor is defined as the ratio of the effective temperature of the spot region to the effective temperature of the surrounding stellar surface.} of the spot. From the point of view of the fit quality, there is obviously some degeneracy between a dim-factor larger or less than 1.0 and the longitude of the spot. 
Finally, we note that there is a degeneracy between the size of the spot and the dim-factor: to the first order, bigger spots with dim-factors closer to unity yield the same effect on the lightcurve as smaller spots with dim-factors that deviate more from unity.

As a result of the degeneracies discussed above, the spot parameters inferred from the fits are certainly not unique. Nevertheless, including such a spot substantially improves the fit quality (see Fig.\,\ref{propsols2} and Table\,\ref{table2}). The only exception is  option I, where the improvement is marginal  because the high value of the $\chi^2$ in this case mostly reflects the impact of the red noise component. 

\begin{table*}[t!]
  \caption{Parameters of the best fit with a binary model including a spotted secondary and consistent with $ABL_{\rm DR2}$\label{table2}}
  \begin{center}
  \begin{tabular}{c c c c c c c c}
    \hline
    LC option & $f_1$ & $f_2$ & $i$\,($^{\circ}$) & \multicolumn{3}{c}{Spot} & $\chi^2$ \\
    \cline{5-7}
    & & & & long. ($^{\circ}$) & radius ($^{\circ}$) & dim &\\
    \hline
    I   & 0.646 & 0.980 & 28.4 &  95.6 &  7.6 & 0.65 & 28.3\\
    II  & 0.464 & 0.980 & 27.0 & 212.9 & 22.9 & 1.12 & 27.9\\
    III & 0.646 & 0.980 & 29.2 & 203.4 & 51.9 & 1.03 &  1.6\\
    \hline
  \end{tabular}
  \tablefoot{The corresponding fits are illustrated in Fig.\,\ref{propsols2}. For comparison, the $\chi^2_{\rm min}$ values for the full parameter space that we explored are 28.1, 27.1, and 1.5 for lightcurve options I, II, and III, respectively. Spot longitudes of $90^{\circ}$ and $270^{\circ}$ correspond to the trailing and leading side of the secondary star, respectively.}
  \end{center}
\end{table*}

In addition to the general trends for the fits discussed in Sect.\,\ref{wospot}, Fig.\,\ref{propsols2} reveals that the $\chi^2$ displays an even flatter behaviour between $f_2 = 0.72$ and $f_2 = 1.00$ than in Fig.\,\ref{propsols}. This observation clearly emphasizes the need for additional independent constraints (provided by the {\it Gaia}-DR2 parallax) to overcome this degeneracy.  

Whether we consider the models with or without a spot on the secondary's surface, we reach the same general conclusions for each of the three lightcurve options: in order for the model parameters to agree with the {\it Gaia}-DR2 parallax, we need a secondary filling factor $f_2 \geq 0.96$ along with an orbital inclination $i$ in the range between $23^{\circ}$ and $31^{\circ}$.\\

\section{Discussion\label{discuss}}
\subsection{Brightness ratio}
A puzzling problem that affects the analysis of several evolved massive binaries is the discrepancy between the visual brightness ratio inferred from spectroscopic and photometric analyses. For instance, in the case of LSS~3074, \citet{Raucq2} found a spectroscopic brightness ratio between the primary and secondary close to 2.5, whilst the photometric data favoured a value near 1.1. \citet{Raucq2} tentatively attributed this effect to the impact of strong radiation pressure on the shape of the binary components \citep[see e.g.\ Fig.\ 3 in][]{Palate}. A very similar discrepancy was reported by \citet{Linder} in the case of 29~CMa. 

At first sight, this discrepancy is also present in the case of HD~149\,404. As pointed out above, leaving the Roche lobe filling factors unconstrained in the fits of the orbital part of the lightcurve actually leads to situations where both stars have nearly identical filling factors. Given the mass-ratio and the effective temperatures of the stars, these configurations would correspond to bolometric flux ratios $l_{\rm bol,1}/l_{\rm bol,2} \simeq 3.3$, which in turn implies an optical brightness ratio $l_1/l_2 \simeq 2.1$, significantly at odds with the spectroscopic value of $0.70 \pm 0.12$. To check whether or not a visual brightness ratio of $2.1$ could be consistent with the spectroscopic data, we  performed test calculations with the CMFGEN code to fit the reconstructed spectra of the two components assuming $l_1/l_2 = 2.1$. We could not achieve a good fit of the strength of the spectral lines in the reconstructed spectra. Moreover, this  brightness ratio would also affect the $\log{g}$ of the primary star, which would change from $3.55$ to $3.10$ \citep[error of $0.15$\,dex,][]{Raucq}, leading to further difficulties in reproducing the reconstructed spectra. We thus favoured the solutions of the lightcurve where the filling factors were constrained to be consistent with $l_1/l_2 = 0.70 \pm 0.12$. In our case, the difference in the quality of the fits between constrained and unconstrained values of the filling factors is relatively small. We thus conclude, at least in the case of HD~149\,404, that the photometric lightcurve is consistent with the spectroscopic brightness ratio.

\subsection{Stellar parameters}
The fits of the {\it BRITE} lightcurve combined with independent information from the {\it Gaia}-DR2 lead to inclinations in the range between $23^{\circ}$ and $31^{\circ}$. The confidence range on $i$, however, is  not centred on that of the {\it Gaia} parallax. The lower the inclination $i$, the closer the $ABL$ parameter is to the value corresponding to the {\it Gaia} parallax. While values of $i$ below $23^{\circ}$ would be allowed by the $ABL$ parameter, they are excluded by the fact that $f_2$ should not exceed unity (a contact or overcontact configuration is ruled out  by the spectroscopic brightness ratio and by the shape of the lightcurve). Therefore, whilst it appears likely that the actual value of $i$ is closer to $23^{\circ}$ than to $31^{\circ}$, the currently available information does not allow us to formulate more stringent constraints on this parameter.

The remaining uncertainty on $i$ has important consequences on our knowledge of the stellar masses. For $i = 23^{\circ}$, we obtain $m_1 = (42.2 \pm 3.5)$\,M$_{\odot}$ and $m_2 = (25.5 \pm 2.2)$\,M$_{\odot}$, whilst $i = 31^{\circ}$ instead leads to $m_1 = (18.4 \pm 1.5)$\,M$_{\odot}$ and $m_2 = (11.1 \pm 1.0)$\,M$_{\odot}$. \citet{Rauw} inferred an inclination of $21^{\circ}$ by comparing the minimum dynamical masses of the stars with typical masses of stars of the same spectral type as quoted by \citet{HP}. However, this comparison could be biased. On the one hand, we now know that the present-day secondary star has undergone a Roche lobe overflow episode in the past (as the mass donor) and displays non-solar surface abundances \citep{Raucq}. The mass-luminosity relation of this star could thus strongly deviate from that of genuine O-stars. A similar but less extreme deviation could apply to the primary star (the mass gainer). On the other hand, the typical masses of O-type stars were revised by \citet{Martins}. Therefore, if we instead compare the primary minimum mass with the spectroscopic mass of an O7.5\,I star as quoted by \citet{Martins}, we find that an inclination of $(24.3 \pm 0.7)^{\circ}$ is required for the primary to have a typical mass for its spectral type.

In summary, whilst we cannot rule out the possibility that the components of HD~149\,404 might be overluminous for their masses, we find that by assembling all the currently available information, it seems likely that at least the mass of the primary is relatively close to typical, unlike the situation encountered in the post-RLOF contact binary LSS~3074 \citep{Raucq2}.

\begin{figure*}
\begin{minipage}{8cm}
\includegraphics*[width=\textwidth,angle=0]{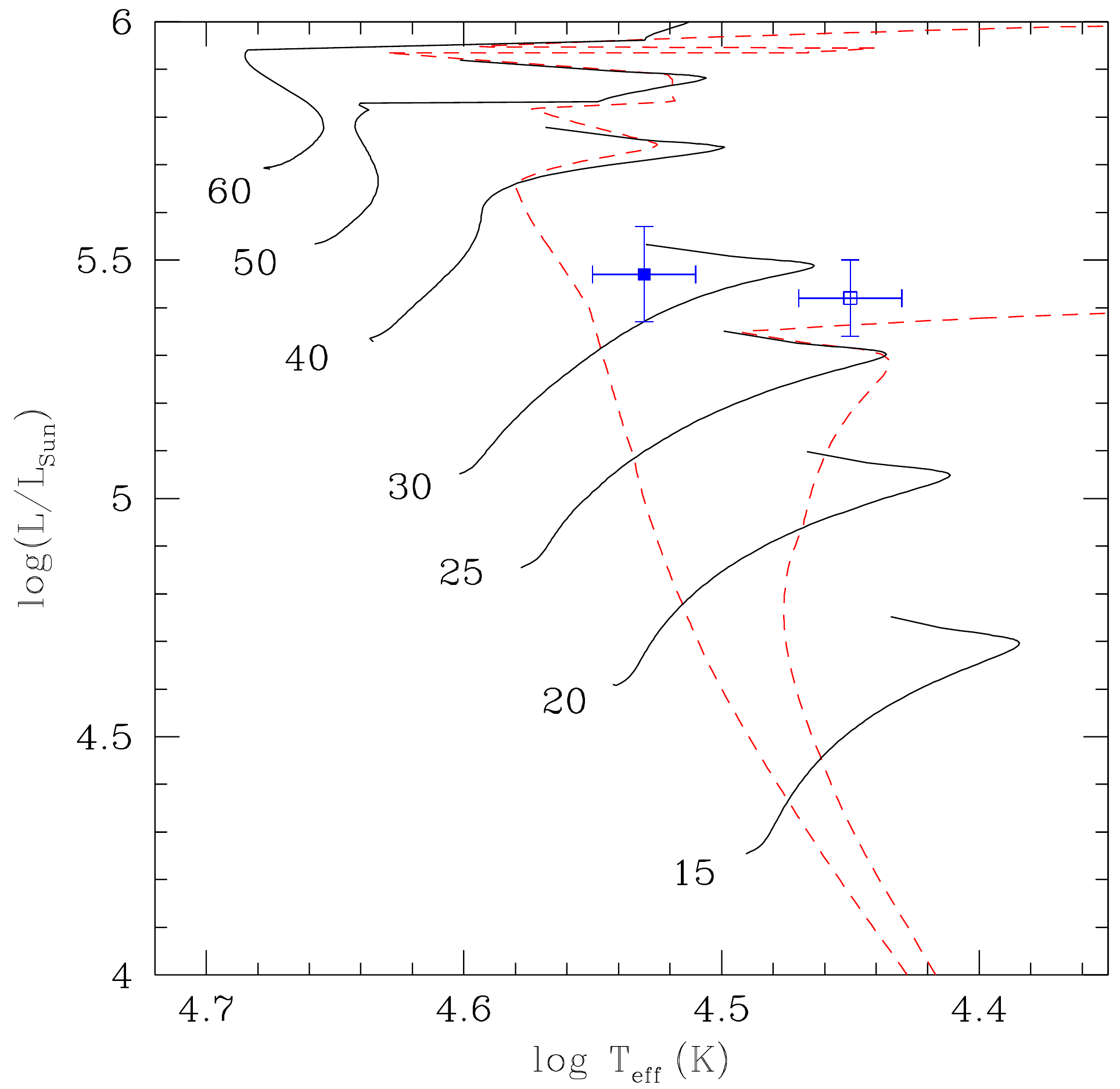}
\end{minipage}
\begin{minipage}{8cm}
\includegraphics*[width=\textwidth,angle=0]{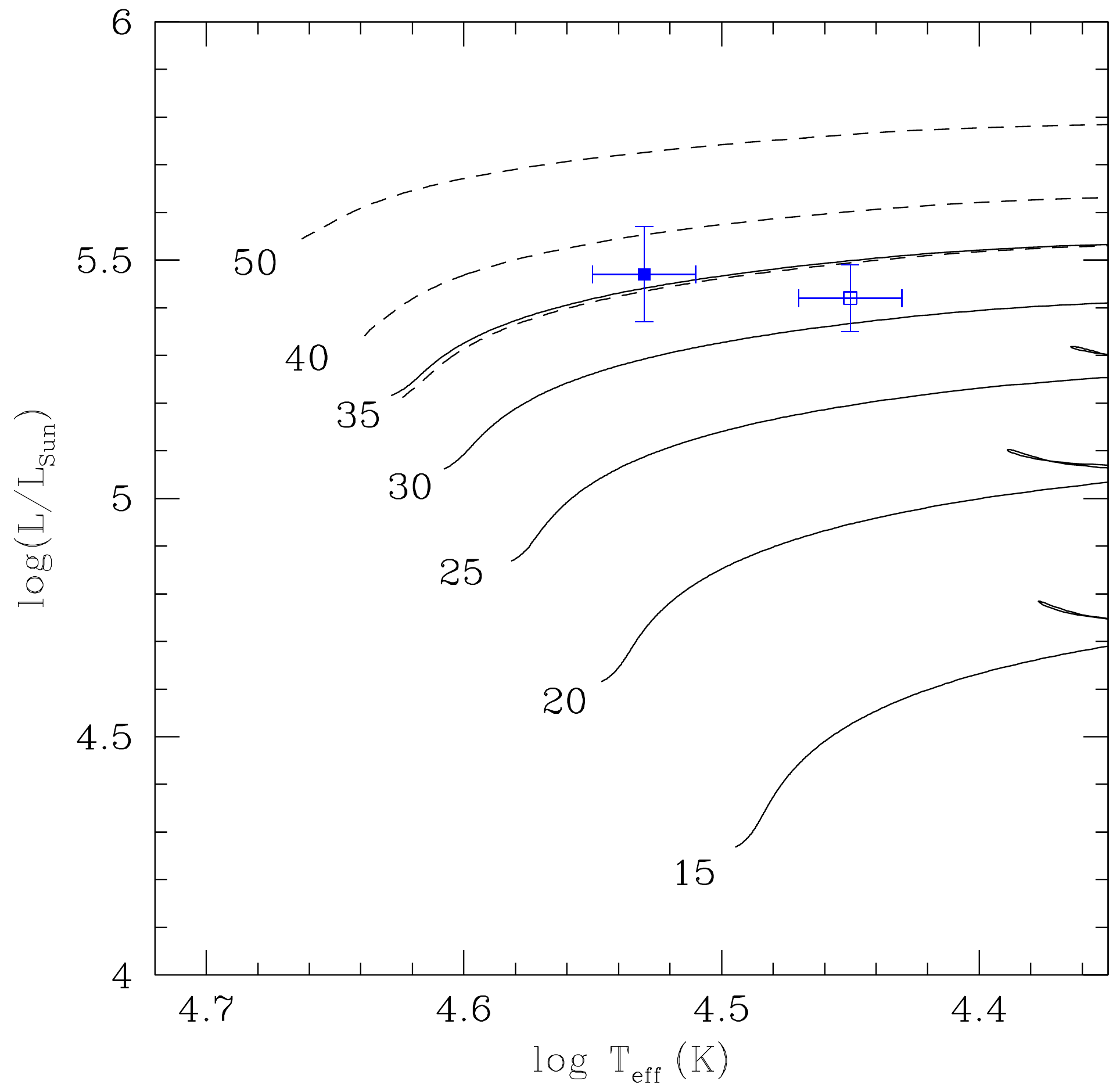}
\end{minipage}

\begin{minipage}{8cm}
\includegraphics*[width=\textwidth,angle=0]{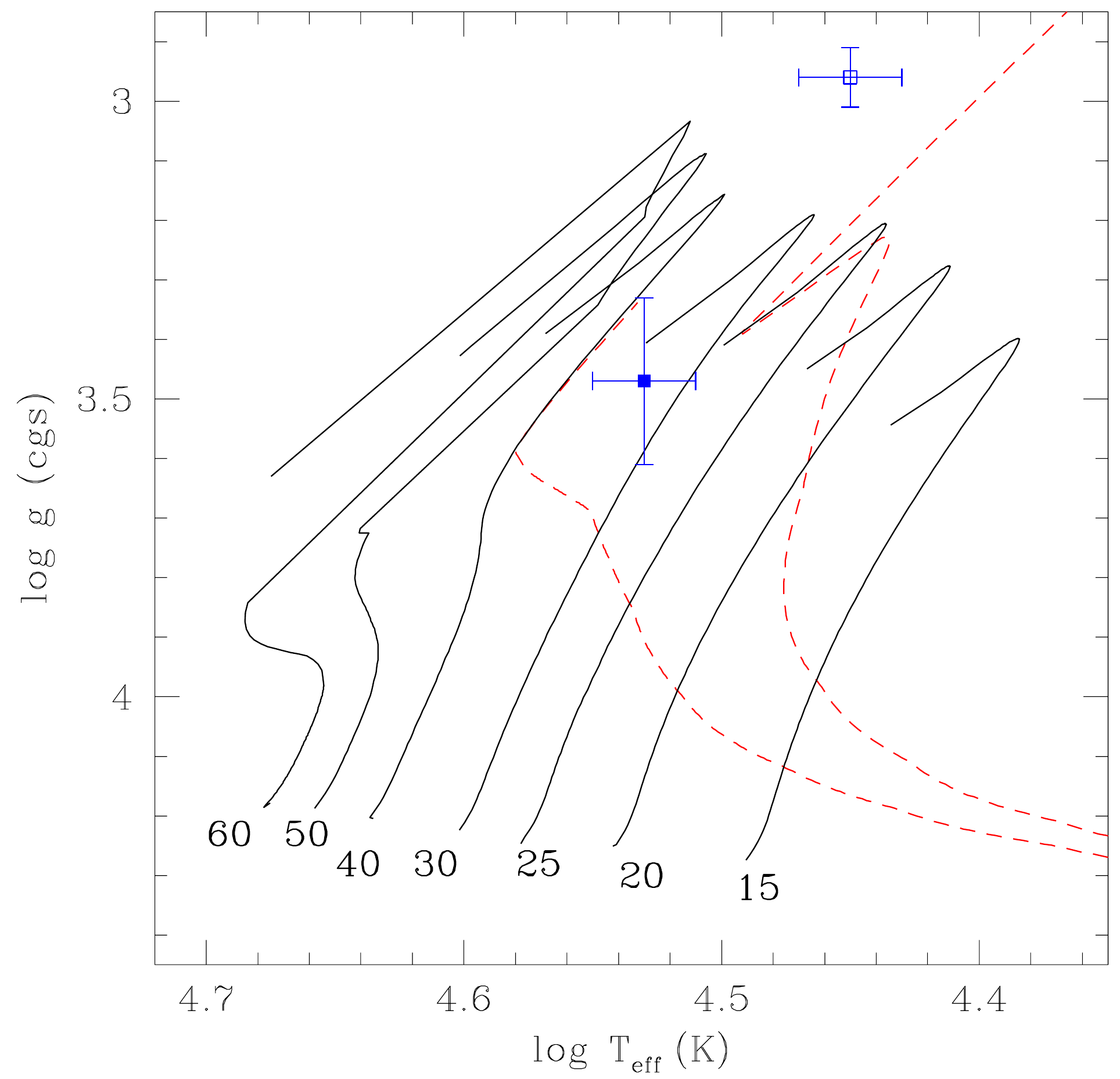}
\end{minipage}
\begin{minipage}{8cm}
\includegraphics*[width=\textwidth,angle=0]{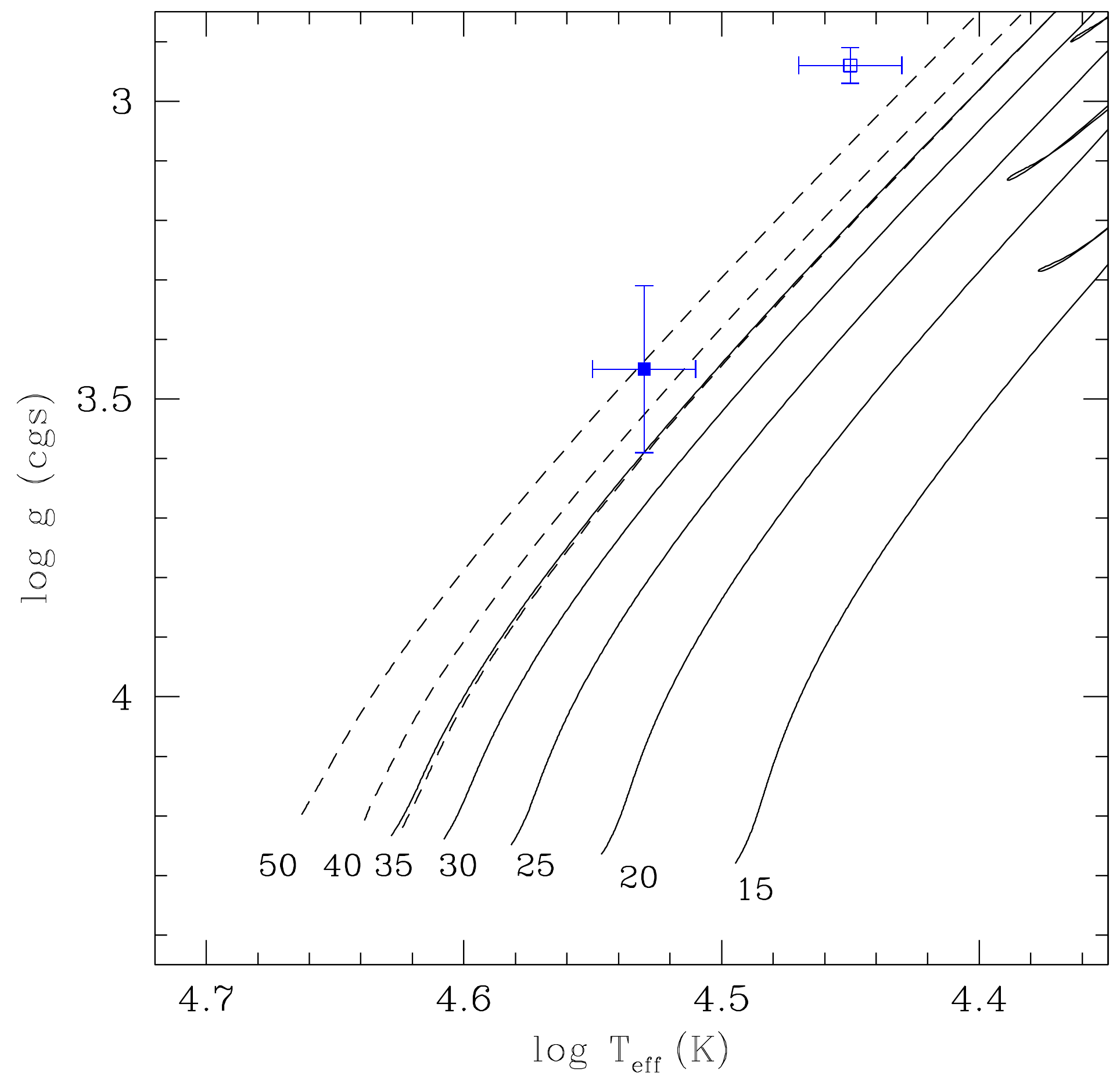}
\end{minipage}
\caption{Position of the components of HD~149\,404 in the Hertzsprung-Russell diagram (upper panels) and in the $\log{g}$ -- $\log{T_{\rm eff}}$ diagram (lower panels). The filled (resp.\ empty) symbols correspond to the primary (resp.\ secondary) star. In the left panels, evolutionary tracks for single massive stars at solar metallicity and initially rotating at $0.4 \times v_{crit}$ from \citet{Ekstrom12} are overplotted. The dashed red lines correspond to isochrones of 5.0 and 7.9\,Myr. In the right panels we show evolutionary tracks from \citet{Brott}. The solid lines yield tracks with initial equatorial rotational velocities near 100\,km\,s$^{-1}$ (112, 111, 110, 109, and 109\,km\,s$^{-1}$ for the 15, 20, 25, 30, and 35\,M$_{\odot}$ models), whilst the dashed tracks correspond to initial equatorial rotational velocities near 200\,km\,s$^{-1}$ (216, 214, and 213\,km\,s$^{-1}$ for the 35, 40, and 50\,M$_{\odot}$ models).
\label{figHRD}}
\end{figure*}

For each lightcurve solution that falls into the acceptable range of $ABL$ values, we compute the associated $\log{g}$ values of both stars. In this way, we obtain $\log{g_1} = 3.45 \pm 0.14$ and $\log{g_2} = 2.94 \pm 0.03$, which are in good agreement with the spectroscopic values of $\log{g_1} = 3.55 \pm 0.15$ and $\log{g_2} = 3.05 \pm 0.15$ \citep{Raucq}. The same models have $\log{L_1/L_{\odot}} = 5.47 \pm 0.10$ and $\log{L_2/L_{\odot}} = 5.42 \pm 0.08$. We can visualize these results in the Hertzsprung-Russell diagram (HRD) and the $\log{g}$--$\log{T_{\rm eff}}$ diagram (see Fig.\,\ref{figHRD}). For qualitative comparison, we also show the evolutionary tracks of \citet{Ekstrom12} with an initial rotational velocity of $0.4 \times v_{crit}$ and of \citet{Brott} with initial rotational velocities that are closest to the $v\,\sin{i}$ values determined by \citet{Raucq} and corrected for $i$ in the range between $23^{\circ}$ and $31^{\circ}$. In addition to different assumptions on the solar metallicity and the evolution of angular momentum, these models also differ in terms of the adopted core overshooting parameter: 0.1 for \citet{Ekstrom12} and 0.335 for \citet{Brott}. Such single-star evolutionary models are certainly not representative of the evolution of the components of HD~149\,404, especially for the present-day secondary which has undergone a RLOF episode;  nonetheless, they can  provide some interesting information. 

For the primary, the position in the HRD falls between single-star evolutionary tracks of 30 -- 40\,M$_{\odot}$, in relatively good agreement with the mass estimates that we have obtained above. This result is supported by the position of the star in the $\log{g}$--$\log{T_{\rm eff}}$ diagram. All in all, the properties of this star seem in fair agreement with the expected value given its current mass. This conclusion is consistent with a scenario where the mass gainer was not totally mixed during the RLOF episode and thus rapidly settled onto an evolutionary track corresponding to its new mass \citep{Vanbeveren98}.

The situation of the secondary is more complex. In the HRD its position suggests a mass between 25 and 30\,M$_{\odot}$ for the \citet{Ekstrom12} tracks or between 30 and 35\,M$_{\odot}$ for the \citet{Brott} models. Not surprisingly, in view of its evolutionary state, the secondary thus appears overluminous for its mass. The secondary is also far away from the tracks corresponding to its mass in the $\log{g}$--$\log{T_{\rm eff}}$ diagram. 

Assuming that the rotational axes of the stars are aligned with their orbital axis, we use the $v\,\sin{i}$ values determined by \citet{Raucq} along with the radii determined for the light curve solutions that fall into the acceptable range of $ABL$ values to assess the rotation rates of both stars. In this way, we find $P_{\rm rot, 1} = (3.98 \pm 0.52)$\,days and $P_{\rm rot, 2} = (8.13 \pm 0.13)$\,days. Our results thus confirm the strong asynchronicity of the rotation of the primary star noted by \citet{Raucq}. This supports the idea that the primary was spun up during the mass and angular momentum transfer. The above rotation periods correspond to roughly 36\% of the critical rotation rates for both stars. This result, along with our discussion of the primary position in the HRD, suggests that the primary was not fully mixed during the mass transfer as it would otherwise display an even faster rotation. It might also imply that mass transfer was more conservative than suggested by the models of \citet{Petrovic}. Finally, we note that the rotation of the secondary is faster than the orbital rate. This situation can be explained if the binary orbital period increased as a result of the RLOF mass exchange as predicted by \citet{Vanbeveren98}.  

\subsection{Wind--wind and binary interactions}
As shown in Sect.\,\ref{modelorb}, including a spot on the surface of the secondary improves the fits of the orbital lightcurve of HD~149\,404. This spot asymmetry could also affect the strength of the spectral lines. Previous studies have noted variations in the equivalent widths (EWs) of the primary and secondary absorption lines with orbital phase \citep{SK,Naze,Rauw}, which were considered  manifestations of the so-called Struve-Sahade effect. In the case of HD~149\,404 these variations concern both the primary and secondary star, which vary in opposite ways. For instance, the equivalent width of the He\,{\sc i} $\lambda$\,4471 line is larger at phase 0.25 than at phase 0.75 for the primary, whilst the reverse situation is seen for the secondary \citep{Naze}. In \citet{Rauw}, we considered the possibility that the variations in the equivalent widths could reflect variations in the continuum at the level of a few percent. Given the amplitude of the variations of the EWs \citep[a factor of two for the He\,{\sc i} $\lambda$\,4471 line,][]{Naze}, this scenario was deemed unlikely, and we instead favoured the possibility of blends with weak emission wings. The results obtained in the present study confirm that the variations in the continuum are indeed rather modest, but they also highlight an aspect which was not considered in the previous work:  the spotted region is not only brighter, but also hotter than the neighbouring parts of the stellar surface. For temperature-sensitive spectral lines, the amplitude of the change in EW can exceed the change in the level of the continuum by a significant factor.   

In Sect.\,\ref{wspot}, we argue that the spot is likely related to the interaction between the stellar winds of HD~149\,404. An alternative could be a spot induced by a strong magnetic activity. However, unless this feature comes from a localized magnetic field, this is rather unlikely to be the case here. Indeed, dedicated spectropolarimetric observations resulted in no detection of a global large-scale magnetic field associated with either the primary or the secondary star \citep{Bfield}. Therefore, interactions between the stars and their wind interaction zone are more likely to be responsible for the spot. Since we show here that the secondary is close to filling its Roche lobe, it seems rather unlikely that its stellar wind can reach a high speed before it collides with the primary's wind. Given the short orbital period, the ratio of  ${\rm v}_{\rm orb}$ to ${\rm v}_{\rm wind}$ must be rather large and the wind interaction zone in HD~149\,404 must therefore be strongly bent by the Coriolis pseudoforces. This situation brings the strongly distorted interaction region very close to the secondary's surface on the leading side of the star and could thus lead to a strong local heating effect.

An alternative explanation for the spot could be an ongoing Roche lobe overflow of the secondary. In such a configuration, the gas stream flowing through the first Lagrangian point L$_1$ is expected to impact the primary star on its rear side \citep{LS}, thereby leading to a hot spot on the rear of the primary (which is most visible around phases 0.75 when the primary is receding from the observer). In Sect.\,\ref{wspot} we assume the spot to be located on the surface of the secondary star. However, there is a degeneracy between a spot on the leading side of one star and a similar spot on the rear side of its companion. This ongoing RLOF scenario would explain the changes of the apparent spectral type of the primary from O6.5 near $\phi = 0.75$ to O7.5 around $\phi = 0.25$ as reported by \citet{Naze}. Whilst we cannot discard this possibility with the currently available data, we do note  that this scenario only works for a subfamily of the models considered here. The ongoing RLOF, and thus the hot spot on the primary star, only exist for a semi-detached configuration with the secondary exactly filling its Roche lobe (i.e.\ $f_2 = 1.0$).

\section{Summary and conclusions\label{conclusion}}
The {\it BRITE-Heweliusz} lightcurve and archival SMEI data of HD~149\,404 reveal two types of photometric variability: a deterministic modulation with the orbital period and a stochastic signal consistent with red noise. The present study mainly focused on the orbital component, which we interpreted as ellipsoidal variations. These variations could be equally well explained by models covering a wide range of values of the filling factors and orbital inclination. However, combining the constraints from the {\it BRITE} and SMEI data with the {\it Gaia}-DR2 parallax, we were able to constrain the secondary filling factor to $f_2 \geq 0.96$ and the orbital inclination to $23^{\circ} \leq i \leq 31^{\circ}$. These constraints lead to a clearer view of the properties of the system and allow us to confirm and refine some of the previous assertions regarding the post-RLOF status of this binary. HD~149\,404 should now be considered as a test case for theoretical binary evolution models: whilst many of its properties qualitatively agree with the expectations for post-RLOF systems, some quantitative aspects probably require further theoretical modelling. 

\begin{acknowledgements}
The operation of {\it BRITE-Heweliusz} is supported by a SPUB grant of the Polish Ministry of Science and Higher Education (MNiSW). This work has made use of results from the ESA space mission {\it Gaia}, the data from which were processed by the {\it Gaia} Data Processing and Analysis Consortium. The Li\`ege team acknowledges financial support through an ARC grant for Concerted Research Actions, financed by the Federation Wallonia-Brussels, from the Fonds de la Recherche Scientifique (FRS/FNRS) through an FRS/FNRS Research Project (T.0100.15), and through two PRODEX contracts (Belspo) related to {\it XMM-Newton} and {\it Gaia} DPAC. GH acknowledges support from the Polish NCN grant 2015/18/A/ST9/00578. APi acknowledges support from the NCN grant 2016/21/B/ST9/01126. AFJM is grateful for financial aid from NSERC (Canada) and FQRNT (Qu\'ebec). APo was supported by NCN grant 2016/21/D/ST9/0065. SMR and GAW acknowledge support from the NSERC (Canada). KZ acknowledges support from the Austrian Space Application Programme (ASAP) of the Austrian Research Promotion Agency (FFG).  
\end{acknowledgements}

\begin{appendix}
\section{Fourier periodogram of the SMEI data\label{appendix1}}
\begin{figure*}[htb!]
\begin{minipage}{8cm}
\begin{center}
\resizebox{8cm}{!}{\includegraphics{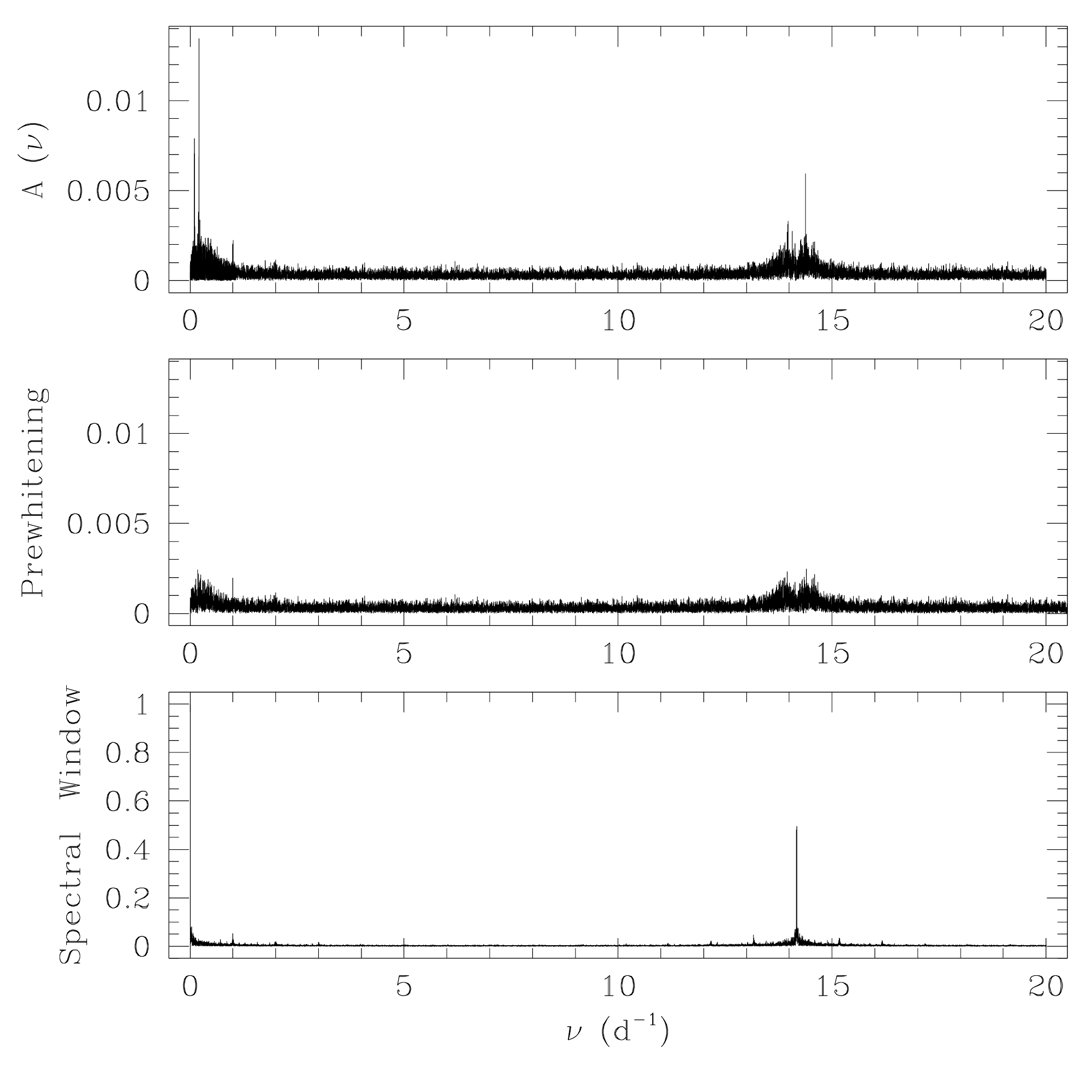}}
\end{center}
\end{minipage}
\hfill
\begin{minipage}{8cm}
\begin{center}
  \resizebox{8cm}{!}{\includegraphics{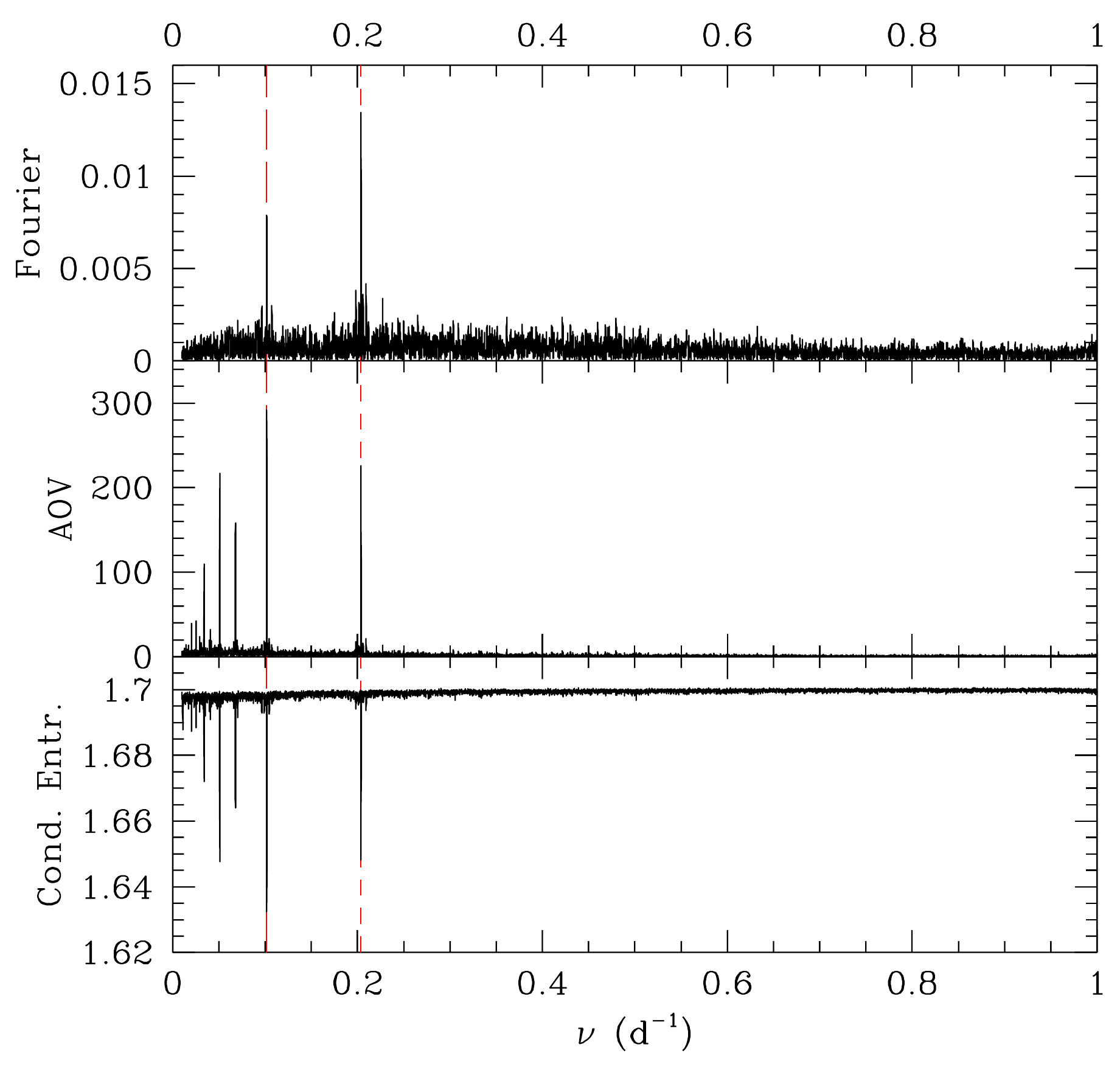}}
\end{center}
\end{minipage}
\caption{Same as Fig.\,\ref{periodogram}, but for the SMEI photometry of HD~149\,404 corrected for the annual modulation, outliers, and long-term trends. The long-dashed (respectively short-dashed) red lines in the right panels correspond to $\nu_{\rm orb}$ (respectively $2 \times \nu_{\rm orb}$).\label{periodogramSMEI1}}
\begin{minipage}{8cm}
\begin{center}
\resizebox{8cm}{!}{\includegraphics{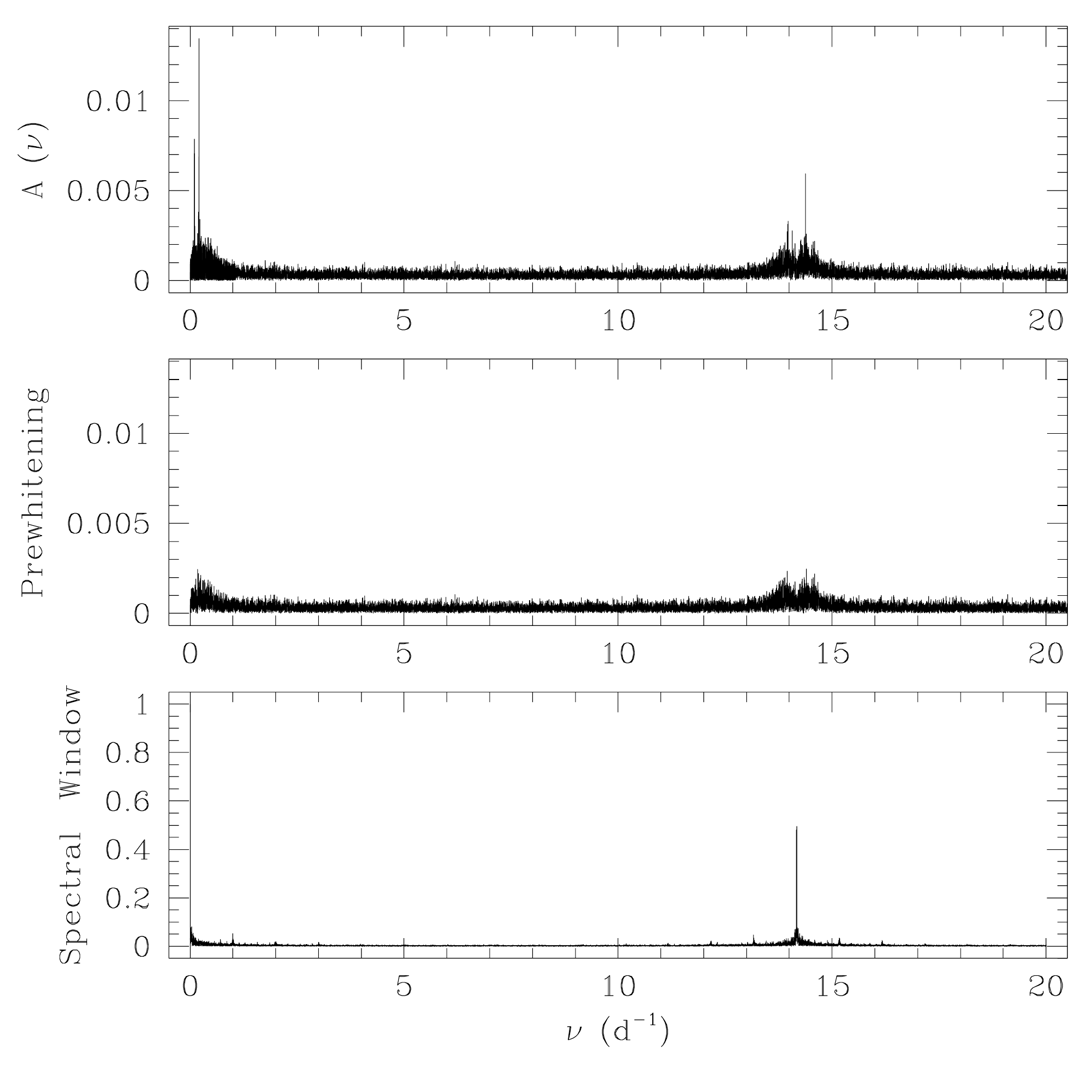}}
\end{center}
\end{minipage}
\hfill
\begin{minipage}{8cm}
\begin{center}
  \resizebox{8cm}{!}{\includegraphics{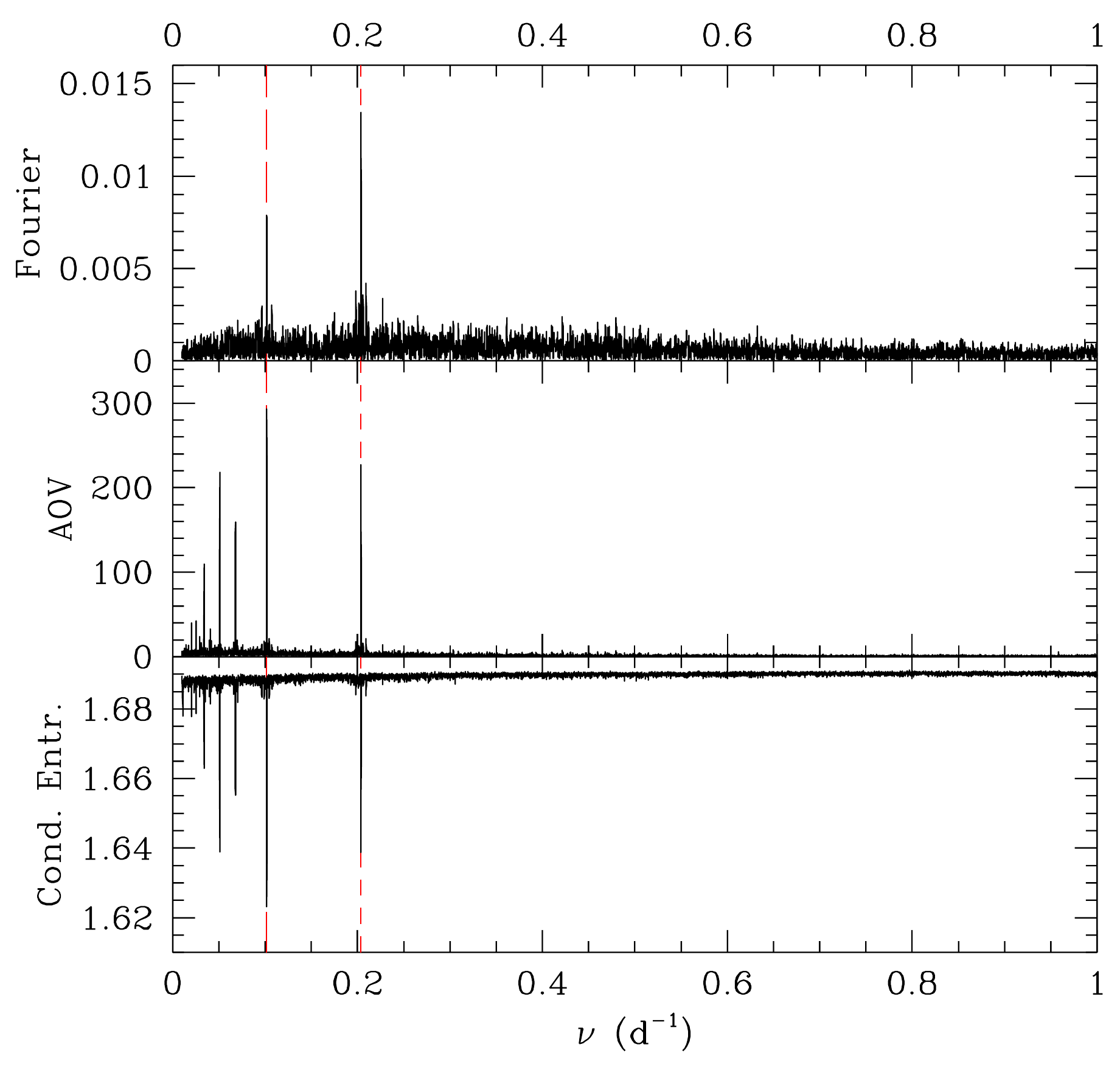}}
\end{center}
\end{minipage}
\caption{Same as Fig.\,\ref{periodogramSMEI1}, but after an additional cleaning step to remove the instrumental signal at frequencies near integer multiples of 1\,d$^{-1}$ and around 3.6\,d$^{-1}$.\label{periodogramSMEI2}}
\end{figure*}

Figures\,\ref{periodogramSMEI1} and \ref{periodogramSMEI2} illustrate the Fourier periodogram of the SMEI time series of HD~149\,404. The entire SMEI data set covers 2448\,days, leading to a natural width of the peaks of $3.51\,10^{-4}$\,d$^{-1}$. In Fig.\,\ref{periodogramSMEI1}, we consider the SMEI time series after correcting for the annual modulation and other long-term trends and removal of clear outliers. The strongest peaks are at frequencies of $(0.203786 \pm 0.000035)$\,d$^{-1}$ and $(0.101912 \pm 0.000035)$\,d$^{-1}$, corresponding respectively to periods of $(4.90711 \pm 0.00085)$ and $(9.81239 \pm 0.00338)$\,days. These values agree within the error bars with half the orbital period and the orbital period as determined from radial velocities \citep[see][and Table\,\ref{param}]{Rauw}. The periodogram further displays some peaks near integer multiples of 1\,d$^{-1}$. 
These features are of instrumental origin. We also note a peak at 0.227\,d$^{-1}$ which could be of astrophysical origin. Given the coarse PSF of the SMEI instrument, we cannot exclude the possibility that this signal arises in another object, unrelated to HD~149\,404. The SIMBAD catalogue lists 145 objects within a radius of 40\arcmin\ around HD~149\,404. The majority of these sources are fainter than 9th magnitude and should thus not affect the SMEI photometry in a significant way. There are, however, three exceptions:  V~1061\,Sco (M1, $V=8.26$) at 15.2\arcmin, HD~149\,340 (A0\,IV, $V = 7.81$) at 21.9\arcmin, and HD~149\,711 ($\equiv$ V~1003\,Sco, B2\,V, $V = 5.83$) at 39.5\arcmin. V~1061\,Sco is classified as a long-period variable and is therefore unlikely to be responsible for the $0.227$\,d$^{-1}$ peak. No information is available about variability of HD~149\,340. HD~149\,711 is classified as an ellipsoidal variable, though no period of variability is available in the literature. The International Variable Star Index (VSX) catalogue does not list any known variable star with a period corresponding to the 0.227\,d$^{-1}$ frequency in the vicinity of HD~149\,404.

In Fig.\,\ref{periodogramSMEI2}, we provide the periodograms after removing the signals at the 1\,d$^{-1}$ (four peaks near that frequency) and 2\,d$^{-1}$ (a single peak) frequencies. Except for a small shift of the peak corresponding to the orbital period, the strongest peaks have the same frequencies as in Fig.\,\ref{periodogramSMEI1}: $(0.203786 \pm 0.000035)$\,d$^{-1}$ and $(0.101911 \pm 0.000035)$\,d$^{-1}$, corresponding respectively to periods of $(4.90711 \pm 0.00085)$ and $(9.81248 \pm 0.00338)$\,days. All these conclusions are backed up by the other period search methods (see the right panels of Figs.\,\ref{periodogramSMEI1} and \ref{periodogramSMEI2}). Using our pre-whitening routine with the frequencies corresponding to $P_{\rm orb}$ (as quoted in Table\,\ref{param}) and its second harmonic, we determine semi-amplitudes of the modulations of $A_1 = 0.0078 \pm 0.0002$\,mag and $A_2 = 0.0135 \pm 0.0002$\,mag.

The periodograms of the SMEI data also display a red noise component. At first sight this looks similar to the red noise seen in the {\it BRITE} data; however,  we note that SMEI data are affected by a red noise of instrumental origin and it is not possible to quantify the relative importance of the two contributions to the red noise seen in this periodogram. 

\end{appendix}

\end{document}